\documentclass{article}
\topmargin = - 0.5 cm
\usepackage{amsmath}
\usepackage{amssymb}
\newcommand{\epr}{{\sc epr}}
 \newcommand{\beq}{\begin{equation}}
\newcommand{\eeq}{\end{equation}}
\newcommand{\bea}{\begin{eqnarray}}
\newcommand{\eea}{\end{eqnarray}}

 \newcommand{\qm}{quantum
mechanics} 
\newcommand{\ca}{$C^*$-algebra}

\newcommand{\Hs}{Hilbert space}

 \newcommand{\ovl}{\overline}
 
\newcommand{\raw}{\rightarrow}

\newcommand{\ot}{\otimes}

\newcommand{\Tr}{\mbox{\rm Tr}\,}

 \newcommand{\BH}{\mathcal{B}({\mathcal H})}

\newcommand{\er}{\eqref}
 \newcommand{\varep}{\varepsilon}

\newcommand{\rh}{\rho} \newcommand{\sg}{\sigma}

 \newcommand{\ps}{\psi} \newcommand{\Ps}{\Psi}
\newcommand{\om}{\omega} 

\newcommand{\CA}{{\mathcal A}} \newcommand{\CB}{{\mathcal B}}

 \renewcommand{\H}{{\mathcal H}}

\renewcommand{\L}{\label}

\newcommand{\C}{{\mathbb C}} 
 
 \newcommand{\R}{{\mathbb R}}
 
  \makeatletter
\newskip\tempskip \def\endproof{{\parfillskip24\p@ plus\@ne
fil\@@par}\tempskip\prevdepth
\ifdim\lastskip=\z@\tempskip\z@\else\vskip-\lastskip
\ifdim\tempskip>4\p@ \tempskip.5\tempskip \else \tempskip\z@\fi\fi
\nobreak\vskip-\baselineskip\vskip-\tempskip\noindent\hbox
to\hsize{\hfill
$\blacksquare$}\par\vskip\tempskip\vskip\abovedisplayskip\@doendpe}
\makeatother \makeatletter
\newskip\tempskip \def\endiproof{{\parfillskip24\p@ plus\@ne
fil\@@par}\tempskip\prevdepth
\ifdim\lastskip=\z@\tempskip\z@\else\vskip-\lastskip
\ifdim\tempskip>4\p@ \tempskip.5\tempskip \else \tempskip\z@\fi\fi
\nobreak\vskip-\baselineskip\vskip-\tempskip\noindent\hbox
to\hsize{\hfill
$\Box$}\par\vskip\tempskip\vskip\abovedisplayskip\@doendpe}
\makeatother 

\begin{document}
\setlength{\baselineskip}{1\baselineskip}
\thispagestyle{empty}
\title{When champions meet: \\ Rethinking the Bohr--Einstein debate}
\author{\textbf{N.P. Landsman} \\ \mbox{} \hfill \\
Radboud Universiteit Nijmegen\\
Institute for Mathematics, Astrophysics, and Particle Physics\\
Toernooiveld 1, 6525 ED NIJMEGEN\\
THE NETHERLANDS\\
\mbox{} \hfill \\
email \texttt{landsman@math.ru.nl}}
\date{}
\maketitle
\begin{abstract}
 Einstein's philosophy of physics (as clarified by Fine, Howard, and Held) was predicated on
his {\it Trennungsprinzip}, a combination  of separability and locality, without which he believed
objectification, and thereby  ``physical thought" and ``physical laws",  to be impossible.  Bohr's philosophy (as elucidated by Hooker, Scheibe, Folse, Howard, Held, and others), on the other hand, was grounded in a seemingly different doctrine about  the possibility of objective knowledge, namely the necessity of classical concepts.
In fact, it follows from Raggio's Theorem in algebraic quantum theory that -
within an appropriate class of physical theories - suitable mathematical translations of the doctrines
of Bohr and Einstein are equivalent. Thus  - upon our specific formalization -
\qm\ accommodates Einstein's {\it Trennungsprinzip} if and only if it is interpreted {\it \`{a} la} Bohr through classical physics.
Unfortunately, the protagonists themselves failed to discuss their differences in this constructive way,
since their debate was dominated by Einstein's ingenious but ultimately flawed attempts to establish the ``incompleteness" of \qm.
 This aspect of their  debate may still be understood and appreciated, however, as  reflecting a much deeper and insurmountable disagreement  between Bohr and Einstein about the knowability of Nature. Using the theological controversy on the knowability of God as a analogy, we can say that Einstein was a Spinozist, whereas Bohr could be said to be on the side of Maimonides. Thus Einstein's off-the-cuff characterization of Bohr as a `Talmudic philosopher' was spot-on.
\end{abstract}
\begin{center} Keywords: Bohr-Einstein debate, EPR, objectification, completeness of quantum mechanics, Raggio's Theorem
\end{center}
\newpage
  \section{Introduction}
What was the Bohr--Einstein debate about, and who ``won" it? So many commentators
(including the protagonists themselves),\footnote{See primarily Bohr (1949) and Einstein (1949a,b). Pertinent correspondence is discussed and/or contained in  Bohr (1996), Einstein \&\ Born (1969), Fine (1986, 2004), Howard (1985), and Held (1998) - Fine and Howard are the main sources for the important letters exchanged by Einstein and Schr\"{o}dinger, and Held contains the most detailed discussion of them.}
 so many opinions. To set the stage, here are a few, to be read and compared pairwise (subtlety increasing in descending order):
  \begin{quote}
 `In fact, in his  first part of his life when he did his really important work, his notion of simplicity were [sic] the guide to the 20th century insofar as science is concerned. Later on I think he was just completely off base. I mean if Einstein had stopped doing physics in the year 1925 and had gone fishing, he would be just as beloved, just as great. It would not have made a damn bit of difference.' (Pais, 1991, in a TV-documentary on Einstein (Kroehling, 1991).)
  \end{quote}
   \begin{quote}
   `During this clarification process [of \qm] Einstein was the first to raise certain issues that still occupy physicists and philosophers - such as the separability of spatially distant systems, or, even more importantly, the measurement problem. These problems, however, were merely stepping stones towards a more fundamental critique:
  Einstein eventually unearthed a conflict between \qm\ and seemingly unavoidable
  common sense opinions on physical reality.' (Held, 1998, p.\ 72.)\footnote{Translated from the German original  by the present author.}
  \end{quote}
 \begin{quote}
`I am now ready to state why I consider Bohr to be not only a major figure in physics but also one of the most important twentieth-century philosophers. As such he must be considered the successor to Kant (\ldots)' (Pais, 2000, p.\ 23.)
 \end{quote}
\begin{quote}
`Now, one can read almost anything into these intriguing asides, from Plato to Wittgenstein. They reveal Bohr's philosophical hang-ups, no more. The careful phraseology of complementarity, drawing on this reservoir, endows an unacceptable theory of measurement
with mystery and apparent profundity, where clarity would reveal an unsolved problem.'
(Bub, 1974, pp.\ 45--46.)
\end{quote}
\begin{quote}
`The refutation of Einstein's criticism does not add any new element to the conception of complementarity, but it is of great importance in laying bare a very deep-lying opposition between Bohr's general philosophical attitude and the still widespread habits of thought belonging to a glorious but irrevocably bygone age in the evolution of science.'
(Rosenfeld, 1967, p.\ 129.)
\end{quote}
\begin{quote}
`It becomes clear how provisional Einstein not only regarded the physics of his time but especially also its epistemological assessment with which we are concerned here.' (Scheibe, 2001, p.\ 126.)
\end{quote}
\begin{quote}
`It is crucial to understand at the outset that Einstein's specific objections to quantum theory did not aim at anything so physically superficial as attempting to show a formal inconsistency in quantum theory. They were aimed, rather, at exposing an inability on the part of the theory to give an adequate account of physical reality. They are, thus, primarily physical, metaphysical, and epistemological in nature, however much they may employ the formal mathematical technicalities of quantum theory. To miss this drive in the objections is not only to fail to understand them; it is to miss the relevance of Bohr's reply and the importance of the ensuing debate.' (Hooker, 1972, p.\ 69.)
\end{quote}
\begin{quote}
`We find that by the Spring of 1927 Einstein had already arrived at the following lines of criticism of the newly emerging quantum theory: (1) the equations of the theory are not relativistically invariant; (2) it does not yield the classical behaviour of macroscopic objects to a good approximation; (3) it leads to correlations among spatially separate objects that appear to violate action-by-contact principles; (4) it is an essentially statistical theory that seems incapable even of {\it describing} the behavior of individual systems; and (5) the scope of the commutation relations may not in fact be so broad as the theory supposes. (\ldots) I believe that these initial disagreements were the ones that lasted.' (Fine, 1986, p.\ 28.)
\end{quote}

What are we to make of this? There is no doubt that, after decades of derision by the Copenhagen camp,\footnote{Perhaps less so by Bohr himself than by his allies. See the archetypal quotations of Pais and Rosenfeld above and of Pauli in Section \ref{einstein} below, and note also the intellectual portrait Pais (1982) paints of the later Einstein.}
Einstein's star as a critic of \qm\  has been on the rise since about the early 1980s.
In the philosophy of physics literature, Howard (1985) and Fine (1986) were
signs of the time, while around the same time  Einstein, Podolsky, \&\ Rosen (1935) began
a second life so as to become one of the most influential papers in twentieth-century physics (see Section \ref{einstein} below). Thus  theoretical and even experimental physicists came to value  Einstein's later contributions to quantum theory almost as much as his earlier ones.

 Bohr's reputation as an interpreter of \qm\  seems to be travelling  in the opposite direction. During his lifetime, Bohr was revered like a demi-god by many of his contemporaries,\footnote{Cf.\ Wheeler (1985, p.\ 226): `Nothing has done more to convince me that there once existed friends of mankind with the human wisdom of Confucius and Buddha, Jesus and Pericles, Erasmus and Lincoln, than walks and talks under the beech trees of Klampenborg Forest with Niels Bohr.' See also hagiographical volumes such as French \&\ Kennedy (1985) and Pais (1991).} certainly because of his brilliant pioneering work on quantum theory, probably also in view of the position of  inspirer and even father-figure
he held with respect to  Pauli (who seems to have been Bohr's greatest admirer) and especially Heisenberg, and  perhaps also to some extent because he `brainwashed a whole generation of theorists into thinking that the job [of giving an adequate philosophical presentation of quantum mechanics] was done fifty years ago' (Gell-Mann, 1979, p.\ 29).
The road for utterances like this had been prepared by physicists such as Bohm, Bell, \&\ Bub,\footnote{See Bell (1987, 2001) and Cushing (1994) for this development.}  but  Bohr-bashing became blatantly bellicose with Beller (1999).  Although even  authors sympathetic to Bohr had previously complained
about his obscurity and idiosyncracy,\footnote{`Bohr's mode of expression and manner of argument are individualistic sometimes to the point of being repellent (\ldots) Anyone who makes a serious study of Bohr's interpretation of \qm\ can easily be brought to the brink of despair' (Scheibe, 1973).} Beller went further than any critic  before or after her by portraying Bohr not as the Gandhi of 20th century physics (as in Pais,  1991) but rather as its Stalin, a philosophical dilettante who knew no mathematics and hardly even followed the physics of his day, but who  nonetheless managed to stifle all opposition
 by a combination of political manoeuvring, shrewd rhetoric, and
 spellbinding  both his colleagues and the general audience by the allegedly unfathomable depth of his thoughts (which according to Beller were actually incoherent and inconsistent).

Despite Beller's meticulous research and passionate arguments, we do not actually believe Bohr's philosophy of \qm\ was such a great muddle after all. Although Beller (1999)
deserves high praise for her courage, and is surely right in criticizing  Bohr for his portrayal of his  doctrine of classical concepts and the ensuing complementarity interpretation of \qm\
 as  absolute {\it necessities} instead of  as the intriguing {\it possibilities} which they really are,\footnote{This point had earlier been made in a  less aggressive manner by - among others, probably - Scheibe (1973, Ch.\ I; 2001, \S VI.27)
 and Cushing (1994). } and also in her analysis of the many obscurities if not inconsistencies of Bohr's early (i.e.\ pre-1935) philosophical thought on \qm\ (see also Held, 1998),
 she goes much too far in denying the coherence and depth  of Bohr's mature (i.e.\ post-1935) philosophy of quantum theory. By the same ``Great Law of the Pendulum",\footnote{An expression used to describe British politics, which tends to swing from Labour to Tory Governments and back, each in turn holding an excessive majority in Parliament.} Beller (1999) as well as Howard (2004a) at first quite rightly draw attention to the fact that the so-called ``Copenhagen Interpretation" is not really the coherent doctrine on \qm\ jointly formulated by Bohr, Heisenberg, and Pauli around 1927 it is traditionally  supposed to be.\footnote{See also  Hooker (1972), Scheibe (1973), and Hendry (1984), where a similar point is made in a friendlier way.} But they subsequently fail to report that Bohr and Heisenberg in fact  came to agree on many basic aspects of the interpretation of \qm, especially
 on the doctrine of classical concepts and its practical implementation by the ``Heisenberg cut" (Scheibe, 1973;  Camilleri, 2005). Indeed, wherever Bohr is ambiguous or hard to interpret for other reasons, finding a reading that agrees with the mature Heisenberg (1958) is a safe way of arriving at a coherent interpretation of \qm. See Section \ref{bohr} below.

 Where many presentations of the Bohr--Einstein  debate (e.g., Rosenfeld, 1967; Folse, 1985; Murdoch, 1987; Whitaker, 1996) closely follow Bohr (1949), we quite agree with Beller (1999) that Bohr's account was written from a winner's perspective, concentrating on parts of the debate where he indeed
emerged victorious, if not ``triumphant".\footnote{Though Bohr `only rejoiced in victory if in winning it he had also deepened his own insight into the problem' (Rosenfeld, 1967, p.\ 131).} Apart from Bohr's own presentation in 1949,  Ehrenfest's widely known  letter of 3 November 1927 to his associates Goudsmit, Uhlenbeck, and Dieke at Leiden undoubtedly also played a role in this perceived outcome of the Bohr--Einstein debate:
\begin{quote}
`Brussels--Solvay was fine! Lorentz, Planck, Einstein, Bohr, Heisenberg, Kramers, Pauli, Dirac, Schr\"{o}dinger, De Broglie (\ldots) and I.
{\sc Bohr} towering completely over everybody. At first not understood at all (\ldots), then step by step defeating everybody. Naturally, once again the awful Bohr incantation terminology. Impossible for anybody else to summarize. (Poor Lorentz as interpreter between the British and the French who were absolutely unable to understand each other. Summarizing Bohr. And Bohr responding with polite despair.) (Every night at 1 a.m.\ Bohr came into my room just to say {\sc one single word} to me, until three a.m.) It was delightful for me to be present during the conversations between Bohr and Einstein. Like a game of chess. Einstein all the time with new examples. In a certain sense a perpetuum mobile of the second kind to break the {\sc uncertainty relation}. Bohr from out of philosophical smoke clouds constantly searching for the tools to crush one example after the other. Einstein like a jack-in-the-box: jumping out fresh every morning. Oh, that was priceless. But I am almost without reservation pro Bohr and contra Einstein. His attitude to Bohr is now exactly like the attitude of the defenders of absolute simultaneity towards him. (\ldots)
!!!!!!! {\sc bravo Bohr} !!!!!!' (Ehrenfest to Goudsmit et al., 1927.)\footnote{See Bohr (1985), pp.\ 415--418 for the German original and ibid.\ pp.\ 37--41 for the English translation.}
\end{quote}

Among supporters of Bohr and of Einstein alike, the general opinion has prevailed  that the central theme of the Bohr--Einstein debate was the (in)completeness of \qm,\footnote{See practically   all older literature, as well as the recent (and insightful) discussions of  De Muynck (2004) and Whitaker (2004).}   the early phase of the debate consisting of Einstein's attempts to  debunk Heisenberg's uncertainty relations (and Bohr's refutations thereof), the later phase - following Einstein's acceptance of the
uncertainty relations - being dominated by Einstein's attacking the alleged completeness of \qm\ {\it despite} the validity of these relations.  Now, there is no doubt that the (in)completeness of \qm\ {\it was} of great  importance to  Bohr and Einstein, and that although they ended up locked in a stalemate themselves,
their discussions of this theme were incredibly fruitful and informative for later developments in the foundations of \qm. For example, Einstein's arguments directly inspired  Schr\"{o}dinger's cat (Fine, 1986; Held, 1998), introduced  what are now called delayed-choice experiments (cf.\ Auletta (2001) for a survey) and, last but not least,  they led to  \epr\ (on whose exceptional importance see below).
Finally, with the exception  of his controversial reply to \epr, Bohr's refutations of Einstein's arguments {\it were} extremely thoughtful and elegant.

There was, however, another side to the debate, where a common battleground not only existed, but could even have led to a reconciliation of the opinions of our great  protagonists.  Namely,  as pointed out  by Held (1998, Ch.\ 6), Bohr and Einstein were both quite worried about the problem of {\it objectification} in physics, especially in \qm. Indeed, since both were thoroughly familiar with the field of epistemology as it had developed since Kant, this problem played a predominant role in their philosophical thought. As reviewed in Sections \ref{bohr} and \ref{einstein} below,  Bohr and Einstein
  were by no means naively anti-realist or realist, respectively, and partly for this reason one might hope to find convergence of their views on this matter. At first sight,
 Bohr and Einstein addressed the problem of objectification in seemingly very different ways:
{\it  \begin{itemize}
\item Bohr claimed objectification of a quantum system through the specification of an experimental context;\footnote{Bohr saw the issue of objectification in classical physics as unproblematic, see Section \ref{bohr}.}
\item Einstein claimed objectification of any physical system to arise from its (spatial) separation from the observer.
\end{itemize}}

Despite appearances, however, only  two steps divide us from a complete identification of these solutions:
\begin{enumerate}
\item The specification of an {\it experimental} context has to be replaced by a specification of a {\it classical} context;
\item The two solutions have to be translated into mathematical language.
\end{enumerate}
Both points are entirely unproblematic; the first is explicit in Bohr's own writings (see Section \ref{bohr}), and the second can be performed with the aid of algebraic quantum theory (cf.\ Section \ref{raggiothm}).
 Having done this, we show that a theorem of Raggio (1981, 1988) yields equivalence of Bohr' solution of the problem of objectification in quantum theory with Einstein's.\footnote{ Attempts to gain some  equivalence between any aspect of the thought of Bohr and Einstein
  are troubled by an opinion that is widely held - probably also  by Bohr and Einstein themselves - to the effect that Einstein's arguments were put forward as requirements on what Nature has to be like,  whereas  Bohr's position (at least in his later period) concerned the linguistic rules of physics (i.e.\ how we think and talk about nature). For example:
  `However, we wish to emphasize that Bohr is not so much concerned with what is {\it truly} real for the distant system as he is with the question of what we would be {\it warranted in asserting} about the distant system from the standpoint of classical description.' (Halvorson \&\ Clifton, 2002). See also Honner (1987). Seen in this way, our protagonists appear to be irreconcilable. We do not share this opinion, but concede that in translating  the positions of Bohr and Einstein  into mathematical criteria
 we have gained common mathematical ground at the expense of some of  the philosophical luggage.  It is up to the reader to decide whether this approach bears any fruit - the author think it does. In any case, we will recover a different philosophical parcel that the author believes to be at the heart of the Bohr-Einstein debate in our closing section; see Section \ref{maimonides} below.}

On this note, the layout of this paper is as follows.
In Section \ref{bohr} we try to clarify those parts of Bohr's philosophy of physics that are relevant to a comparison of his position with Einstein's. This mainly refers to Bohr's doctrine of classical concepts, as Einstein never really entered into a discussion of the principle of complementarity.\footnote{`The sharp formulation of which, moreover,
 I have been unable to achieve despite much effort which I have expended on it.' (Einstein, 1949b, p.\ 674). See also Held (1994, 1998). } Here we combine what we feel to be the clearest passages in Bohr's own writings with some of the interpretations of commentators such as Hooker (1972), Scheibe (1973), Folse (1985), and Howard (1994). Subsequently, in Section \ref{einstein} we do the same for Einstein, closely following Howard (1985) and Fine (1986), with additional insights from Held (1998).
This leads to the identification of Einstein's {\it Trennungsprinzip} (separability principle)
as the cornerstone of his doctrine.
Although it is clear from the work of these authors (also cf.\ Deltete \&\ Guy, 1991) that  \epr\ was really a confused and confusing mixture of  Einstein's earlier attack on the uncertainty relations  with his later
``incompleteness" arguments against \qm\ (not to speak of the smokescreen
erected by Bohr's reply),\footnote{As Schr\"{o}dinger put it in a letter to Einstein
dated July 13, 1935:
`It is as if one person said, ``It is bitter cold in Chicago"; and another answered, ``That is a fallacy, it is very hot in Florida".' (Fine, 1986, p.\ 74).}
 we still comment on this paper.
This is  partly  because  the immediate response to \epr\ by the Bohr camp reveals their breathtaking arrogance  towards Einstein's critique of quantum theory, but more importantly, because what are now quite rightly called
 \epr-correlations form an essential part of modern physics.
 For example, the whole field of quantum cryptography hinges on them,
as does the  associated phenomenon of quantum teleportation (surely one of the most spectacular predictions of quantum theory, now duly verified in the lab).
Amazingly, the {\it one} outcome of the Bohr--Einstein debate that is of lasting value for physics therefore concerns a phenomenon whose {\it existence} Einstein actually denied
(as he used \epr-correlations in  a {\it reductio ad absurdum} argument),
and whose {\it significance} Bohr utterly failed to recognize!

In Section \ref{meet} we create an imaginary, conciliatory ``Bohr",  who - perhaps even less realistically! -
happens to be familiar with algebraic quantum theory.\footnote{For introductory accounts see  Primas (1983),  Emch (1984),  or Haag (1992). In 1953--54  Rudolf Haag
(one of the pioneers of algebraic quantum theory) was a postdoc at Copenhagen
in the {\sc cern} theory group led by Bohr!} Our ``Bohr" realizes that (at least in a world where physical observables are represented by operators on a Hilbert space)\footnote{This incorporates the possibility of a classical world as well as of a quantum one.}
an appropriate mathematical translation of his doctrine of classical concepts is equivalent to an analogous formalization of Einstein's {\it Trennungsprinzip},   applied to the measuring instrument in combination with the measured quantum system. As already mentioned, this equivalence follows from a theorem of Raggio (1981, 1988), and our application of it to the Bohr--Einstein debate owes a great deal to both Primas (1983) and Held (1998). We explain this theorem in Section \ref{raggiothm}. 

Seen  through mathematical glasses (and hence dropping some of the ideology), the positions  of our two  giants therefore overlapped significantly  - a point both failed to recognize, probably not merely  for the ideological reason stated above, but undoubtedly also because of the desire of both to defeat the opponent.
Taking this unfortunate desire for granted, who actually won the debate? Folk wisdom has it that Bohr did, but in Section \ref{whowon} we argue on the basis of our analysis that {\it on the terms of the debate} it was in fact Einstein who should have emerged as the victor! 

More importantly, the  {\it agreement} between Einstein and Bohr on the solution to the problem of objectification in quantum theory paves the way for an identification of their exact {\it disagreement} on the issue of the (in)completeness of this theory. Namely,
 the technical parts of their debate on the (in)completeness of \qm\ just served as a pale reflection of  a much deeper philosophical disagreement between Bohr and Einstein about the knowability of Nature. For Bohr's doctrine of classical concepts implies that no {\it direct} access to the quantum world is possible, leaving its essence unknowable. This implication was keenly felt by Einstein, who in response was led to characterize his opponent  as a `Talmudic philosopher'. In the last section of this paper we
 try to show how astute this characterization  was through a theological analogy, in which  Bohr and Einstein on the (un)knowability of Nature are compared with Maimonides and Spinoza on the (un)knowability of God, respectively. Although there is no evidence that Bohr was  familiar with the work of Maimonides (Spinoza's influence on Einstein, instead, is well documented),  at least the author has been greatly enlightened by the comparison. We hope the reader is, too.
\section{Bohr's doctrine}\label{bohr}
Protestantism is based on the idea that everything worth knowing about religion is written in the Bible.  Taking the Dutch Republic as an example, within the general {\it Protestant Church} one had the  Calvinist {\it Dutch Reformed Church},  within which
disagreements about the interpretation of the Bible (here specifically concerning Predestination) eventually became so heated that the political leader of the so-called {\it Remonstrants} (who believed in some degree of Free Will), State Pensionary Johan van Oldenbarnevelt, was beheaded in 1619 on the orders of the figurehead of the {\it Contra-Remonstrants} (as the enemies of the Remonstrants were aptly called), Prince Maurits of Orange. This conflict tore apart and debilitated  Dutch society for almost a century  (Israel, 1995).
Similarly, Trotskyism is predicated on the notion that {\it the} political understanding of the world and {\it the} right course of action to be taken to improve it can be found in the works of Leon Trotsky. Typically, however, `Trotskyist parties and groups are notorious for their tendency to split into smaller groups, quarrelling over theoretical differences that seem insignificant or indecipherable to an outsider, but which sometimes have major practical consequences for those who hold those positions.'\footnote{See \texttt{http://encyclopedia.laborlawtalk.com/Trotskyism}, as well as \texttt{http://www.broadleft.org/trotskyi.htm}
 for a list of  international umbrella Trotskyist organizations that exist as of July 2005 (competing largely with each other, rather than with their alleged joint enemy, world capitalism).}

Thus one is intrigued by the suggestion of Howard (1994) - made in the light of the undeniable fact that Bohr is often misrepresented and misunderstood -
`to return to Bohr's own words,\footnote{The principal  primary sources are Bohr's Como Lecture, his reply to \epr\, and his essay dedicated to Einstein (Bohr, 1927,  1935, 1949). These papers were actually written in collaboration with Pauli, Rosenfeld, and Pais, respectively.  Historical discussions of the emergence and reception of these papers are given in Bohr (1985, 1996) and in Mehra \&\ Rechenberg (2001).  See also  Bohr (1934) and Bohr (1958), as well as Bohr (1987) for a collection of his philosophical writings chosen by Bohr himself.}
 filtered through no preconceived philosophical dogmas.' Well! Perhaps  Bohr's own words themselves were  responsible for the confusion?
\begin{quote}
`However eminent the abilities of the late Niels Bohr, he certainly did not study the art of writing in such a style, that not only he {\it might possibly be understood} by those of his readers who comprehended the subject nearly as well as himself, but that he {\it could not possibly be misunderstood} by any one of ordinary capacity and attention - an invaluable art (\ldots)' (Wood, 1954, p.\ 98.)\footnote{Well\ldots we have subsituted `(Niels) Bohr' for `Dr Young'.}
\end{quote}

 Indeed, the result is as expected: as Howard himself points out to his credit,  `Bohr's own words' have led Folse (1985) to claim Bohr was a realist, Faye (1991) to portray him as an anti-realist, and  Murdoch (1987) to position him as a neo-Kantian: a possibility, we take the liberty to add, Scheibe (1973)
is conspicuously  silent about, despite his intimate familiarity with it through his mentor C.-F. von Weizs\"{a}cker, who himself claimed Bohr was a Kantian.
 And here we have restricted ourselves to some of the most reliable and illuminating commentators on Bohr  - a group that definitely includes Howard himself, as well as Hooker (1972) and Held (1998).\footnote{It is abundantly clear by now that renowned philosophers of science like  Popper and Bunge completely failed to understand Bohr (Hooker, 1972; Peres, 2002).}

Considerable progress can be made, however, if one relies on Bohr's own words {\it and} on intelligent commentaries on them, such as those  written  by the authors just mentioned. But, as should be clear from the previous paragraph, even this is not enough to arrive at an unambiguous interpretation of Bohr. As a final criterion we therefore propose that it is a good sign when Bohr and Heisenberg agree about a particular notion. Hence complementarity in the sense Bohr meant it is out (Camilleri, 2005), as is Bohr's obscure and obsolete ``quantum postulate",\footnote{\label{QP} The Como Lecture (Bohr, 1927) was entitled `The quantum postulate and the recent development of atomic theory'.  There Bohr stated its contents as follows:  `The essence of quantum theory is the quantum postulate: every atomic process has an essential discreteness -  completely foreign to classical theories - characterized by Planck's quantum of action.' (Instead of `discreteness', Bohr alternatively used the words  `discontinuity' or `individuality' as well. He rarely omitted amplifications like `essential'.) Even more emphatically, in his reply to \epr\ (Bohr, 1935):
`Indeed the finite interaction between object and measuring agencies conditioned by the very existence of the  quantum of action entails - because of the impossibility of controlling the reaction of the object on the measurement instruments if these are to serve their purpose - the necessity of a final renunication of the classical ideal of causality and a radical revision of our attitude towards the problem of physical reality.'} but - and this is in any case the crucial part in Bohr's philosophy as far as it is relevant to his debate with Einstein - the {\it doctrine of classical concepts} is in.\footnote{For Heisenberg's eventual endorsement see Heisenberg (1958) and Camilleri (2005).}
It might be appropriate to quote Bohr's statement of this doctrine from his paper dedicated to Einstein:
 \begin{quote}
 `However far the phenomena transcend the scope of classical physical explanation, the account of all evidence must be expressed in classical terms. (\ldots) The argument is simply that by the word {\it experiment} we refer to a situation where we can tell others what we have done and what we have learned and that, therefore, the account of the experimental arrangements and of the results of the observations must be expressed in unambiguous language with suitable application of the terminology of classical physics.'
  (Bohr, 1949, p.\ 209.)
 \end{quote}

 Our first comment is that the argument is not simple at all; although people like Heisenberg and Pauli must have learned it from Bohr in person,\footnote{
`To me it has not been all that frustrating to follow Bohr's thinking by reading these papers [i.e.\ those contained in Bohr (1987)], an undertaking which does demand care and patience. I realize, however, my uncommon advantage of many discussions with Bohr about his philosophical ideas.' (Pais, 1991, p.\ 422).} less fortunate folk like the present author have to extract it from Bohr's later writings (e.g., the last five essays in Bohr (1958)) and from intelligent commentaries thereon.\footnote{On the origin of the doctrine of classical concepts we especially recommend Hooker (1972), Folse (1985), and Howard (1994).}
The point then turns out to be this:  For Bohr, the \textit{defining} property of classical physics was the property that it was \textit{objective}, in that it could be studied in an observer-independent way:\begin{quote}
`All description of experiences so far has been based on the assumption, already inherent in ordinary conventions of language, that it is possible to distinguish sharply between the behaviour of objects and the means of observation. This assumption is not only fully justified by everyday experience, {\it but even constitutes the whole basis of classical physics.}'
(Bohr, 1958, p.\ 25; italics added.)\footnote{Bohr often regarded certain other properties as essential to classical physics, such as  determinism, the combined use of space-time concepts and dynamical conservation laws, and the possibility of pictorial descriptions. However, these properties were in some sense secondary,  as Bohr considered them to be  {\it consequences} of the possibility of isolating an object in classical physics. For example: `The assumption underlying the ideal of causality [is] that the behaviour of the object is uniquely determined, quite independently of whether it is observed or not' (Bohr, 1937), and then again, now negatively: `the renunciation of the ideal of causality [in \qm] is founded logically only on our not being any longer in a position to speak of the autonomous behaviour of a physical object' (Bohr, 1937). See Scheibe (1973).}
\end{quote}
 Heisenberg
shared this view:\footnote{As Camilleri (2005, p.\ 161) states: `For Heisenberg,
classical physics is the fullest expression of the ideal of objectivity.'}
 \begin{quote}
 `In classical physics science started from the belief - or should one say from the illusion? - that we could describe the world or at least part of the world without any reference to ourselves. This is actually possible to a large extent. We know that the city of London exists whether we see it or not. It may be said that classical physics is just that idealization in which we can speak about parts of the world without any reference to ourselves. Its success has led to the general idea of an objective description of the world.'  (Heisenberg, 1958, p.\ 55.)
 \end{quote}

It is precisely the objectivity of classical physics in this sense that guarantees the possibility of what Bohr calls `unambiguous communication' between observing subjects, provided this communication  is performed `in classical terms'. For if the method of communication is separate from the communicating subjects, it is independent of them, and hence ``objective" in the sense used by Heisenberg in the above quotation - some readers might prefer to equate Bohr's ``unambiguous" with ``intersubjective" instead of ``objective". See also Hooker (1991).

So far, so good. Now,  on the basis of his ``quantum postulate" (see footnote \ref{QP}), Bohr came to believe that in quantum physics the mutual independence of subject (or observer) and object no longer applied. Although
 authors sympathetic to Bohr tend to be remarkably silent about the absence of
Bohr's ``quantum postulate" from any modern axiomatic treatment of
\qm, or even from any serious account of quantum theory that uses its mathematical formalism, one can follow Bohr's argument at this point if one replaces his ``quantum postulate" by the property of entanglement.\footnote{This replacement is  implicit or explicit in practically all of Howard's writings about Bohr, but the present author doubts whether Bohr ever understood the notion of entanglement in the way we do. For we are not talking about an intuitive  pictorial notion of inseparability between two quantum systems caused by the exchange of quanta whose size
infuriatingly  refuses to go to zero, but about an unvisualizable  property of the mathematical formalism of \qm.} In any case, Bohr felt this lack of independence to be a threat to the objectivity of physics. He responded to this threat with a highly original move, namely by still insisting on  the objectivity
- or ``unambiguity"  -  of at least our {\it description} of physics. This objectivity, then, Bohr claimed to be accomplished by `expressing the account of all evidence in classical terms'. Moreover, he insisted that
such an account was {\it necessary} for this purpose.\footnote{
Clearly, if Bohr's previous analysis is correct, it would follow that it is {\it sufficient} that this procedure be followed. As already mentioned in the Introduction, we follow Beller (1999) in holding that whenever it suited his reasoning, Bohr replaced possibility with necessity, rarely if ever giving noncircular arguments for such  replacements.}

Bohr's  apparently  paradoxical doctrine of classical concepts has radical and fascinating consequences. For one, it clearly precludes a completely quantum-mechanical description of the world;  Bohr even considered it pointless to ascribe a state to a quantum-mechanical object considered on its own.
At the same time, Bohr's doctrine  precludes a purely classical description of the world, for underneath classical physics one has quantum theory.  The solution to this dilemma that Bohr and Heisenberg proposed is to divide the system whose description is sought into two parts: one, the object, is to be described quantum-mechanically, whereas the other, the apparatus, is treated \textit{as if it were classical}.
 Thus  the division between object and subject coincides with the one between a quantum-mechanical and a classical description; both divisions are purely epistemological
and have no counterpart in ontology.\footnote{Here we
 take leave from Howard's (1994) analysis; it is frightening that Bohr seems to leave room for such differing implementations of his doctrine of classical concepts
 as Heisenberg's (which we follow, siding e.g.\ with Scheibe (1973)) and Howard's own.  Howard's arguments against Heisenberg's implementation
 are that it `introduces a new dualism into our ontology' and that `one would like to think that the classical/quantum distinction corresponds to an objective feature of the world'. The first has just been dealt with; the second takes a 100 page article to answer (Landsman, 2005).}
In the literature, the division in question is often called the \textit{Heisenberg cut}.
  Despite innumerable claims to the contrary  (e.g.,  to the effect that Bohr held that a separate realm of Nature was intrinsically classical), there is no doubt that both Bohr and Heisenberg believed in the fundamental and universal nature of \qm, and, once more,   saw the classical description of the apparatus as a {\it purely epistemological move},
which expressed the fact that a given {\it quantum} system is {\it being used} as a measuring device.\footnote{In the writings of Bohr terms like `observer', `subject',  `apparatus', `measuring device', `experimental conditions', etc.\ are used interchangeably. Bohr never endorsed a subjective interpretation of \qm, let alone one in which the mind of a human observer plays a role. Quite to the contrary, because of his doctrine of classical concepts, the apparatus acts as a classical buffer between the quantum world and the human observer, so that
Bohr could consistently claim that the problem of observation (in the sense of the human perception of sense data and the like) was of a purely classical nature even in quantum physics. See, e.g., Scheibe (1973) and Murdoch (1987). } Indeed, some of Bohr's most ingenious arguments against Einstein's early attempts to invalidate the uncertainty relations are based on the typical change of perspective in which a system initially used as a classical measuring device is suddenly seen as a quantum system subject to the uncertainty relations (thereby, of course,  surrendering its role as an apparatus).  See Bohr (1949) and Scheibe (1972).

 The idea, then, is that {\it a quantum-mechanical object is studied exclusively through its influence on an apparatus that is described classically}. Although {\it described} classically, the apparatus is supposed to be  influenced by its {\it quantum-mechanical} coupling to the underlying object. A key point in this doctrine  is that  {\it probabilities arise solely  because we look at the quantum world through classical glasses}:
\begin{quote}
`Just the necessity of accounting for the function of the measuring agencies on classical lines excludes in principle in proper quantum phenomena an accurate control of the reaction of the measuring instruments on the atomic objects.'  (Bohr, 1956, p.\ 87.)
\end{quote}
\begin{quote} `One may call these uncertainties objective, in that they are simply a consequence of the fact that we describe the experiment in terms of classical physics; they do not depend in detail on the observer. One may call them subjective, in that they reflect our incomplete knowledge of the world.' (Heisenberg, 1958, pp.\ 53--54.)
\end{quote}
 Hence the probabilistic nature of quantum theory is not intrinsic but extrinsic, and as such is wholly explained by the doctrine of classical concepts, at least conceptually. We feel this to be a very strong argument in favour of Bohr's doctrine. Mathematically, the simplest illustration of this idea is as follows. Take a finite-dimensional Hilbert space $\H=\C^n$ with the ensuing algebra of observables $\CA=M_n(\C)$ (i.e.\ the $n\times n$ matrices). A unit vector $\Ps\in\C^n$ determines a quantum-mechanical state in the usual way. Now describe this quantum system as if it were classical by ignoring all observables except the diagonal matrices. The state then immediately collapses to a probability measure on the set of $n$ points, with probabilities given by the Born rule
$p(i)=|(e_i,\Ps)|^2$, where $(e_i)_{i=1,\ldots,n}$ is the standard basis of $\C^n$. Similarly, the Born--Pauli rule for the probabilistic interpretation of the wave function $\Ps\in L^2(\R^3)$ in terms of $|\Psi(x)|^2$
immediately follows if one ignores all observables on $L^2(\R^3)$ except the position operator.\footnote{Technically, one restricts $\Ps$ - seen as a state on the \ca\ $\CB(L^2(\R^3))$ as explained in Section \ref{raggiothm} -  to the \ca\ $C_0(\R^3)$ given by all multiplication operators on  $L^2(\R^3)$ defined by continuous functions of $x\in\R^3$ that vanish at infinity. This restriction yields a probability measure on $\R^3$, which is precisely the usual one originally proposed by Pauli.}

In a realistic situation, the procedure of extracting a classical description of a quantum system is vastly more complicated, involving the construction  of semiclassical observables
through either macroscopic averaging or taking the $\hbar\raw 0$ limit, which only asymptotically (i.e.\  as the system becomes infinitely large or as $\hbar\raw 0$) form a commutative algebra (Landsman, 2005). The mathematical procedures necessary for a classical description of a quantum system confirm a conceptual point often made by Bohr and Heisenberg, namely that a classical description is always an idealization. Hence the identification  of classical physics with an objective description explained above then implies
that Bohr's ideal of unambiguous communication can only be satisfied in an approximate sense.
 \section{Einstein's doctrine}\label{einstein}
As mentioned in the Introduction, one cannot simply say that Bohr was an anti-realist, and at least since {\it The Shaky Game} (Fine, 1986) straightforward  remarks to the effect that Einstein was a realist would immediately disqualify their author as well.  For   repeated utterances like:
\begin{quote}
`The belief in an external world independent of the perceiving subject is the basis of all natural science.' (Einstein, 1954, p.\ 266.)\footnote{\label{live} The German original is:
`Der Glaube an eine vom wahrnehmenden Subjekt unabh\"{a}ngige Au\ss enwelt liegt aller Naturwissenschaft zugrunde.' (Einstein, 1982, p.\ 159.) It seems to have been an emotional need for Einstein to detach himself from his fellow humans in order to devote himself to the study of the Cosmos. For example, Einstein's former associate  Adriaan Fokker wrote in his highly perceptive obituary of Einstein:
`His true passion was to penetrate the riddle of the immeasurable cosmos, which stood
high above the muddle and the confusion of personal interests, feelings and low impulses of men. Such thought comforted him when he had seen through the hypocrisy
of the common ideals of decency. The consideration of this external reality lured him as a liberation from an earthly prison.'  (Fokker, 1955; translated from the Dutch original by the present author).
Einstein  made a similar point himself: `I mercifully
 belong to those people who are granted  as well as able  to dedicate their best efforts to the consideration and the research of objective, time-independent matters.
How fortunate  I am that this mercy, which makes one quite independent of personal fate and of the behaviour of one's fellow humans,  has befallen me.' (Einstein, 1930.)  Perhaps it should be investigated to what extent this need stood behind Einstein's  insistence on the observer-independence of any physical theory (and \qm\ in particular).
}\end{quote}
are counterbalanced by occasional subtle epistemological analyses like the one given in Einstein (1936). Einstein's approach towards realism is well summarize by his own words:
\begin{quote}
`I did not grow up in the Kantian tradition, but came to understand the truly valuable which is to be found in this doctrine (\ldots). It is contained in the sentence: ``The real is not given to us, but put to us ({\it aufgegeben}) (by way of a riddle)."\footnote{This idea is better expressed in German: ``Das Wirkliche ist uns nicht gegeben, sondern aufgegeben
(nach Art eines R\"{a}tsels)." Here `aufgegeben' had better been translated by `assigned' rather than by `put'.
} This obviously means: there is such a thing as a conceptual reconstruction for the grasping of the inter-personal, the authority of which lies purely in its validation. This conceptual construction  refers precisely to the ``real" (by definition), and every further question concerning the ``nature of the real" appears empty.' (Einstein, 1949b, p.\ 680.)
\end{quote}

In fact, as pointed out by Held (1998, Ch.\ 6),  Einstein's epistemological position was by no means inconsistent with Bohr's - as we shall see, the way Einstein addressed the problem of objectification  was even equivalent (in a suitable mathematical sense) to Bohr's approach (as reviewed in  the preceding section).

As a brief summary,\footnote{See Fine (1986), Held (1998) and Howard (2004b, 2005) for detailed  expositions of Einstein's philosophy of science. See also Deltete \&\ Guy (1991).}  one might say that (the mature) Einstein held that realism was something like a physical postulate, according to which empirical data are supposed to be produced by real objects - which, unlike the empirical data that act as an intermediate between object and observer, are independent of the observer. But which among all the possible kinds of objects that one might conceive  as potential sources of empirical data are real? Einstein's answer to this question, and thereby his solution to the problem of objectification, was that `spatial separation is a sufficient condition for the individuation of physical systems' (Howard, 2004b, \S 5). `[Einstein's]
realism is thus the thesis of spatial separability.' (ibid.).
 The following quotation
is pertinent:
\begin{quote}
`It is characteristic of these physical things [i.e.\ bodies, fields, etc.] that they are conceived of as being arranged in a space-time continuum. Further, it appears to be essential for this arrangement of the things introduced in physics  that, at a specific time, these things claim an existence independent of one another, insofar as these things ``lie in different parts of space". Without such an assumption of the mutually independent existence (the ``being-thus") of spatially distant things, an assumption which originates in everyday thought, physical thought in the sense familiar to us would not be possible. Nor does one see how physical laws could be formulated and tested without such a clean separation. (\ldots)

For the relative independence of spatially distant things $A$ and $B$, this idea is characteristic: an external influence on $A$ has no {\it immediate} effect on $B$; this is known as the principle of ``local action", which is applied consistently only in field theory.
The complete suspension of this basis principle would make impossible the idea of the existence of (quasi-) closed systems and, thereby, the establishment of empirically testable laws in the sense familar to us.'\\ (Einstein, 1948, pp.\ 321--22. Translation by Howard (1985), pp.\ 187--88.)
\end{quote}
And similarly, in a letter to Born: \begin{quote}
`However, if one renounces the assumption that what is present in different parts of space has an independent, real existence, then I do not at all see what physics is supposed to describe. For what is thought to be a ``system" is, after all, just conventional, and I do not see how one is supposed to divide up the world objectively so that one can make statements about the parts.'
(Einstein \&\ Born, 1969, p.\ 223-24. Translation by Howard (1985), p.\ 191.)
\end{quote}
As Howard (1985, p.\ 191)  comments, `what Einstein suggests here is that the separability principle is necessary because it provides the only imaginable objective principle for the individuation of physical systems.' See also Held (1998, Ch.\ 6) for a detailed analysis of Einstein's views on objectification. Here the ``separability principle",
which Einstein called the {\it Trennungsprinzip}, means, according to Howard (1985, p.\ 173) that `spatially separated systems possess separate real states';
in addition, Einstein invokes a ``locality principle", according to which `the state of a system can be changed only by local effects, effects propagated with finite, subluminal velocities' (ibid., p.\ 173).\footnote{There is a subtle difference at this point between Howard (1985) and Fine (1986), which is irrelevant for our story.} And similarly: `[Einstein] thought, not unreasonably, that spatial separation was the only way of distinguishing systems'
(Deltete \&\ Guy, 1991,  p.\ 392).

As clarified by Howard (1985), Fine (1986, 2004), Deltete \&\ Guy (1991), and Held (1998, \S 22), Einstein's doctrine does not come out well in Einstein, Podolsky, \&\ Rosen (1935), abbreviated as \epr\ in what follows.  \epr\ was actually written by Podolsky,\footnote{`For reasons of language [\epr] was written by Podolsky after much discussion. Still, it did not come out as well as I had originally wanted; rather the essential thing was, so to speak, smothered by learnedness.'
Einstein to Schr\"{o}dinger, 19 June, 1935, quoted and translated by Fine (1986), p.\ 35. In particular,
 the most quoted sentence of \epr, viz.\ `If, without in any way disturbing a system, we can predict with certainty the value of a physical quantity, then there exists an element of physical reality corresponding to this physical quantity,' was never repeated or endorsed by Einstein and is now attributed entirely to Podolsky. This did not prevent the speaker in
an {\sc hps} seminar the author once attended from asking the audience to rise while reading it out loud, as it allegedly expressed  Einstein's deepest metaphysical thought.
} and the argument  Einstein himself had in mind was  much simpler than what one finds in \epr.\footnote{See also Howard (1985, 1990).}
 Rosenfeld  recalls:
\begin{quote}
`He [Einstein] had no longer any doubt about the logic of Bohr's argumentation; but he still felt the same uneasiness as before (``Unbehagen" was his word) when confronted with the strange consequences of the theory. ``What would you say of the following situation?" he asked me [following a seminar by Rosenfeld in Brussels in 1933 that Einstein attended].
``Suppose two particles are set in motion towards each other with the same, very large momentum, and that they interact with each other for a very short time when they pass at known positions. Consider now an observer who gets hold of one of the particles, far away from the region of interaction, and measures its momentum; then, from the conditions of the experiment, he will obviously be able to deduce the momentum of the other particle. If, however, he chooses to measure the position of the first particle, he will be able to tell where the other particle is. This is a perfectly correct and straightforward deduction from the principles of \qm; but is it not very paradoxical? How can the final state of the second particle be influenced by a measurement performed on the first, after all physical interaction has ceased between them?"'
 (Rosenfeld, 1967, pp.\ 127--128.)
\end{quote}
The last sentence contains the thrust of the argument, but the reference to two different measurements on the first particle indicates that Einstein was still playing with the idea of undermining the uncertainty relations as late as 1933. In any case, \epr\ is a somewhat incoherent mixture of the latter goal with Einstein's later drive to prove the incompleteness of \qm\ while accepting the uncertainty relations.  In later presentations Einstein omitted any reference to
two ``complementary"  measurements, as exemplified by
 his `Reply to criticisms':
\begin{quote}
`And now just a remark concerning the discussions about the Einstein--Podolski--Rosen Paradox. (\ldots) Of the ``orthodox" quantum theoreticians whose position I know, Niels Bohr's seems to me to come nearest to doing justice to the problem. Translated into my own way of putting it, he argues as follows:

If the partial systems $A$ and $B$ form a total system which is described by its $\psi$-function $\ps/(AB)$, there is no reason why any mutually independent existence (state of reality) should be ascribed to the partial systems $A$ and $B$ viewed separately, {\it not even if the partial systems are spatially separated from each other at the particular time under consideration}. The assertion that, in this latter case, the real situation of $B$ could not be (directly) influenced by any measurement on $A$ is, therefore, within the framework of quantum theory, unfounded and (as the paradox shows) unacceptable.

By this way of looking at the matter it becomes evident that the paradox forces us to relinquish one of the following two assertions:
\begin{trivlist}
\item [(1)] the description by means of the $\ps$-function is {\it complete}
\item[(2)] the real states of spatially separated objects are independent of each other.\footnote{\label{ESfn} Einstein evidently meant this notion of separability to incorporate locality.}
\end{trivlist}

On the other hand, it is possible to adhere to (2), if one regards the $\ps$-function as the description of a (statistical) ensemble of systems (and therefore relinquishes (1)). However, this view blasts the framework of the ``orthodox quantum theory".'
(Einstein, 1949b, pp.\ 681--682.)
\end{quote}
Here  Einstein does not actually paraphrase Bohr very well at all, as Bohr's concept of the ``wholeness" of the system at hand does not merely refer to the indivisibility of the joint system $A \& B$, but to the experimental setup used to primarily  {\it define} and subsequently {\it measure} the classical variables of $A$ and $B$ (such as position and momentum in the \epr\ version and spin in the Bohm version of the thought experiment).\footnote{See all references cited on Bohr so far, and especially Held (1998, Ch.\ 5).}  Furthermore, Einstein's conclusion (as expressed in the penultimate sentence of the above quotation) is highly dubious (Fine, 1986) and, according to the overwhelming majority of physicists,  has been refuted by the work of Bell (1987, 2001) and its aftermath.\footnote{\label{bellit}
A recent and reliable reference is Shimony (2005), and also the older reviews by  Bub (1997) and  Auletta (2001) are still recommended. For an excellent short review see also Werner \&\ Wolf (2001).
 The pertinent papers by Bell himself did not sufficiently clarify the relationship between the separability and locality assumptions of Einstein and the factorizability assumption central to the derivation of the Bell inequalities,
 but this point was  fully elucidated by Jarrett (1984) and others; see  Bub (1997) for the full story this paper initiated, as well as  Scheibe (2001) for an independent resolution.
Moreover, later refinements of Bell's arguments (e.g., Greenberger et al., 1990;  Mermin, 1993; Hardy, 1993) seem to have put an end to what little hope might have been left for the \epr-argument (which is given {\it
within} the theoretical framework of \qm). See Seevinck (2002) for a review.
It now seems that only hardcore determinism
of the kind expressed by 't Hooft (private communication), who denies the freedom  of the experimenter to set the polarization angle of his apparatus, can circumvent Bell's Theorem
(to the effect that ``local realism" is incompatible with \qm).
This stance would probably have resonated well with Einstein himself, who (following Spinoza) denied Free Will (Jammer, 1999). Also cf.\ Section \ref{maimonides}  below.

On the experimental side, certain shortcomings in Aspect-type experiments apparently still leave some room for conspiracy theories that might restore local realism.  See the bibliography in Shimony (2005), supplemented with Hess and Phillip (2001a,b, 2002, 2004, 2005), Seevinck \&\ Uffink (2002), Szabo \&\ Fine (2002), Winsberg \&\ Fine (2003), and Santos (2005) either in favour of local realism or at least against discarding it on the basis of current theoretical and experimental knowledge, and on the other side Grangier (2001),  Gill et al.\ (2002, 2003, 2004), and Myrvold (2003). Noting that
 even Fine (2004) concedes  that more refined future experiments will probably refute local realism for good,
it is hard to avoid the impression that its supporters seem locked in a rearguard fight. In any case, {\it experiments}
 would not save \epr, who argued within \qm, as already remarked.}

Let us therefore concentrate on (1) and (2), whose disjunction in formal quantum theory was a valid inference - if not a brilliant insight -  by Einstein. Unfortunately,
 Einstein (and \epr)  insisted on a further elaboration of this disjunction, namely the idea that there exists some version of \qm\ that {\it is} separable (in the sense of (2)) at the cost of assigning more than one state to a system (two in the simplest case). It is this unholy version of \qm\ that Einstein (and \epr)  called ``incomplete". Now, within the formalism of \qm\ such a multiple assignment of states (except in the trivial sense of wave functions differing by a phase factor) makes no sense at all,  for the entanglement property lying at the root of the non-separability of \qm\  is so deeply entrenched in its formalism  that it simply cannot be separated from it. Largely for this reason, the  stream of papers and books analyzing the ``logical structure" of the \epr-argument  (e.g.,  Krips, 1969; Hooker, 1972; Scheibe, 1973;
McGrath, 1978; Fine, 1986; Redhead, 1987; Deltete \&\ Guy, 1991; Held, 1998; Shimony, 2001; Dickson, 2002) will probably never subside.

As explained in the next section, there
 {\it is} a sense in which quantum theory can be made compatible with Einstein's separability
principle, namely by invoking none other than Bohr; furthermore, there {\it is} a precise sense in which the ensuing version of quantum theory is incomplete in a way Einstein would have recognized in his broader uses of the word, but this has nothing to do with his multiple wave functions. As already pointed out, the ensuing obscurity of \epr\ is further troubled by the absence from it of Einstein's guiding hand.

Nonetheless, whether or not it revealed Einstein's true intentions, and whatever the
quality of its logic, Einstein, Podolsky, \&\ Rosen (1935) is arguably the most famous paper ever written about \qm. For although Einstein's original intention might have been to press what he felt to be a {\it reductio ad absurdum} argument {\it against} quantum theory, the paper is now generally read as stating a spectacular prediction {\it of} quantum theory, viz.\ the existence of what these days are quite rightly called {\it \epr-correlations}.
The theoretical analysis of these correlations and their context by  Bell (1987, 2001) revitalized the foundations of quantum theory, and their experimental verification (in the form of the violation of the Bell inequalities) was done in one of the most stunning series of experiments in twentieth-century  physics (Aspect et al., 1981, 1982a,b, 1992; Tittel et al., 1998).\footnote{See also Baggott (2004) and Shimony (2005) for recent overviews.}  More recently, the avalanche of  papers in which characters called  Alice and Bob appear,
and indeed the whole field of quantum cryptography and large areas of quantum computation, would have been  unthinkable without \epr. Quantum teleportation,  in some sense the ultimate {\it reductio ad absurdum}  prediction of \qm\ inspired by \epr\ (Bennett et al., 1993), has meanwhile moved up from Star Trek to the lab  (Zeilinger, 2000; Ursin et al., 2004).

So what did Bohr and his circle think about \epr?
\begin{quote}
`{\sc Einstein} has once again made a public statement about quantum mechanics, [namely] in the issue of the  Physical Review of May 15 (together with Podolsky and Rosen  - no good company, by the way). As is well known, that is a disaster whenever it happens. ``Because, so he concludes razor-sharply, - nothing can exist if it ought not exist"
(Morgenstern). Still, I must grant him that if a student in one of their earlier semesters had raised such objections, I would have considered him quite intelligent and promising.
(\ldots) Thus it might anyhow be worthwhile if I waste paper and ink in order to formulate those inescapable facts of \qm\ that cause Einstein special mental troubles. He has now reached the level of understanding, where he realizes that two quantities corresponding to non-commuting operators cannot be measured simultaneously and cannot at the same time be ascribed definite numerical values. But the fact that disturbs him in this connection is the way two systems in \qm\ can be coupled to form one single total system.
(\ldots) All in all, those elderly gentlemen like {\sc Laue} and {\sc Einstein} are haunted by the idea that \qm\ is admittedly {\it correct}, but {\it incomplete}.'
(Pauli to Heisenberg, June 15, 1935.)\footnote{The German original
is reprinted in   Bohr (1996), p.\ 480, with English translation on pp.\ 252--253.}
\end{quote}
\begin{quote}
`The small group, Bohr, Heisenberg, Pauli and a few others who through intense debates during many years had become intimately familiar with all aspects of the quantal description, was mainly astonished that Einstein had found it worthwhile to publish this ``paradox" in which they saw nothing but the old problems, resolved long ago, in a new dress.' (Kalckar in Bohr (1996), p.\ 250.)
\end{quote}
\begin{quote}
`The essence of Bohr's reply to Einstein is his demonstration that this new thought experiment does not exhibit {\it any} new features not already inherent in the analysis of the old double-slit experiment debated at the Solvay conference in 1927.' (Kalckar in Bohr (1996), p.\ 255.)
\end{quote}
\begin{quote}
`These remarks apply equally well to the special problem treated by Einstein, Podolsky, Rosen, which has been referred to above, and which does not actually involve any greater intricacies than the simple examples discussed above.' (Bohr, 1935, p.\ 699.)
\end{quote}
\begin{quote}
`It will be seen, however, that we are dealing with problems of just the same kind as those raised by Einstein in previous discussions.' (Bohr, 1949, p.\ 232.)
\end{quote}

As is clearly shown by the above quotations (which could easily be supplemented with many others), Bohr and his allies did not see `any new features'  in \epr, and merely concentrated on the task to `clear up such a misunderstanding at once' (Rosenfeld, 1967, p.\ 128).
Surely, this attitude must count among the most severe errors of judgement in the history of physics. Even so, the ``clearing up" (Bohr, 1935) is done with an obscurity surpassing that of \epr.  Authors sympathetic to (and well-informed about) Bohr are divided on the thrust of his reply (cf.\ the quite different expositions in Folse (1985), Murdoch (1987), and Held (1998)), and even on the question whether or not his reply marks a change in his philosophy of \qm; whereas those hostile to Bohr (Beller \& Fine, 1994; Beller, 1999) even deny its coherence by claiming that Bohr (1935) is an incoherent mixture of his pre-1935 and his post-1935 attitudes (which, then, are claimed to contradict each other). Thus we cannot but agree with detached observers such as Halvorson \&\ Clifton (2002), who claim that  `Although Bohr's reply to the \epr\ argument is supposed to be a watershed moment in the development of his philosophy of quantum theory, it is difficult to find a clear statement of the reply's philosophical point', and Dickson (2002), according to whom `it is notoriously difficult to understand Bohr's reply - over 60 years later, there remains important work to be done understanding it'.\footnote{See Whitaker (2004) for a critical  discussion  of  Halvorson \&\ Clifton (2002) and Dickson (2002).}
\section{Bohr meets Einstein}\label{meet}
What would have been a {\it good} reply to \epr? In a truly successful attempt to ``defeat" Einstein,
 Bohr could have come up with Bell's analysis, whose mathematics even he presumably could have handled. Alternatively, and much more easily, as pointed out by De Muynck (2004) he could have remarked - well within the ``spirit of Copenhagen" - that \epr-correlations are physical only if they are measured, and that measuring them requires operations at both ends (as in the later experiments of Aspect (1981, 1982a,b)), which taken together are nonlocal. This would have countered  the \epr-argument - including their infamous ``elements of reality" - in a decisive way.

 But if, instead,  Bohr had genuinely been interested in finding common ground with Einstein, he could have written him  the following letter:
\begin{quote}
{\it Dear Einstein,

Whether our actual meetings have been of short or long duration, they have always left a deep and lasting impression on my mind, and when writing this I have, so-to-say, been arguing with you all the time. What has always comforted me through the suffering our disagreements - and in particular your rejection of \qm\ and its ensuing complementarity interpretation as a completely rational description of physical phenomena - have caused me, is the joke of two kinds of truth. To the one kind belong statements so simple and clear that the opposite assertion obviously could not be defended. The other kind, the so-called ``deep truths", are statements in which the opposite also contains deep truth. Where we differ, our opposition appears to me to be of the latter sort, as I will now venture to explain. Of course, I am deeply aware of the inefficiency of expression which must make it very difficult to appreciate the trend of the argumentation aiming to bring out the essential ambiguity involved in a reference to physical attributes of objects in dealing with phenomena where no sharp distinction can be made between the behaviour of the objects themselves and the interaction with the measuring instruments. The present account may give a clearer impression of the necessity of a radical revision of basic principles for physical explanation in order to restore logical order in this field of experience.

We have both been grasping for principles that make physics and physical laws, as a human endeavour, possible. We both agree that this very possibility entails - because of the necessity of unambiguous communication between scientists  if these are to serve their purpose - a certain amount of separation between observing subject and observed object,
even though such a separation is a priori denied by the quantum postulate, according to which every atomic process has an essential discreteness -  completely foreign to classical theories - characterized by Planck's quantum of action, in whose elucidation you have played so large a part. Indeed, the new situation in physics has so forcibly reminded us of the old truth that we are both spectators and actors in the great drama of existence which, lacking an author, has no plot. For that very reason, there is no question that not only has the deterministic description of physical events, once regarded as suggestive support of the idea of predestination, lost its unrestricted applicability by the elucidation of the conditions for the rational generalization of classical physics, but it must even be realized that its very failure, and therewith the emergence of the probabilistic kind of argumentation that is, within its proper limits, so characteristic  of the quantum-mechanical description of atomic phenomena,
 lies purely in the necessity of expressing the account of all evidence in classical terms, however far the phenomena transcend the scope of classical physical explanation.

You, however, have equally forcefully urged that
physical thought would be impossible without a spatial separability and locality principle,
in the very sense that spatially separated systems possess separate real states and, moreover, the state of a system can be changed only by effects propagated with subluminal velocities.
From the great experience of meeting you for the first time during a visit to Berlin in 1920 till the present day, the possibility of a  reconciliation of our respective  points of view, so very different as they may appear at first sight,  has been among my greatest hopes. Consequently, the necessity of a renewed examination of the central tenets of our debate has led me to a closer analysis of the issue seemingly dividing us, which has finally brought me to a point of great logical consequence. Namely, with hindsight, the essential lesson of our discussions is that within a large class of theories that incorporates both classical and quantum mechanics,} our very principles coincide.

{\it I remain thus, with cordial greetings, \\ Yours, Niels Bohr}
\end{quote}
\section{Raggio's Theorem}\label{raggiothm}
Our imaginary ``Bohr" here bases his conciliatory gesture  to Einstein on what is sometimes called {\it Raggio's Theorem}
(Raggio, 1981, 1988), which we now briefly explain.\footnote{See also Primas (1983) and  Bacciagaluppi (1993).} This theorem is stated in the language of operator algebras
(Takesaki, 2003), which comes in naturally and handy when discussing the Bohr--Einstein debate, as it enables one to describe classical and quantum theories within the same mathematical framework.\footnote{
See also Landsman (1998), Clifton, Bub, \&\ Halvorson (2003), and Bub (2004) for this strategy.}
Recall that a {\it $C^*$-algebra} is a complex algebra $\CA$ that is
complete in a norm $\|\cdot\|$ that satisfies $\| AB\|\,\leq\, \| A\|\,\|
B\|$ for all $A,B\in\CA$, and has an involution $A\raw A^*$ such that
$\| A^*A\|=\| A\|^2$. A basic example is
$\CA=\BH$, the algebra of all bounded operators on a \Hs\ $\H$, equipped with the usual operator norm and adjoint.   By the Gelfand--Naimark theorem, any \ca\ is isomorphic to a norm-closed self-adjoint subalgebra of $\BH$, for some \Hs\ $\H$. In particular, the algebra $M_n(\C)$ of complex $n\times n$ matrices is a \ca, as is its commutative subalgebra $D_n(\C)$ of diagonal matrices (here $\H=\C^n$ in both cases). Readers unsympathetic towards heavy mathematical formalism may keep these two examples in mind in what follows. The latter is a special case of \ca s of the form
 $\CA=C_0(X)$, the space of
all continuous complex-valued functions on a (locally compact Hausdorff) space $X$ that vanish at infinity,\footnote{In the sense that for every $\varep>0$ there is a {\it compact} subset $K\subset X$ such that $|f(x)|<\varep$ for all $x\notin K$.} equipped with the supremum norm,\footnote{I.e.\
$\| f\|_{\infty}:=\sup_{x\in X} |f(x)|$} and involution given by (pointwise)
complex conjugation. Indeed, one has $D_n(\C)=C_0(\{1,2,\ldots,n\})$ (where the set
$\{1,2,\ldots,n\}$ may be replaced by any set of cardinality $n$).
By the Gelfand--Naimark lemma, any commutative \ca\ is isomorphic to $C_0(X)$ for some locally compact Hausdorff space $X$.

Furthermore, we use the notion of a state that is usual in the operator-algebraic framework. Hence  a {\it state} on a \ca\ $\CA$  is a linear functional
$\rh:\CA\raw\C$ that is {\it positive} in that $\rh(A^*A)\geq 0$ for all $A\in\CA$
and {\it normalized} in that $\rh(1)=1$, where $1$ is the unit element of $\CA$.\footnote{If $\CA$ has no unit one requires that $\|\rh\|=1$.}
Now, if $\CA=M_n(\C)$,  a fundamental theorem of von Neumann states that
each state $\rh$ on $\CA$ is given by a density matrix $\hat{\rh}$ on $\H$, so that $\rh(A)=\Tr (\hat{\rh} A)$ for each $A\in\CA$. In particular, a {\it pure} state on $M_n(\C)$ is necessarily of the form $\ps(A)=(\Ps,A\Ps)$ for some unit vector $\Ps\in\C^n$.\footnote{If $\CA$ is a  von Neumann algebra, a so-called
 {\it normal} state on $\CA$ satisfies a certain  additional continuity condition. If $\CA=\CB(\H)$ for an infinite-dimensional \Hs\ $\H$, then von Neumann's theorem just mentioned says in its full glory that
 each {\it normal} state $\rh$ on $\CA$ is given by a density matrix on $\H$ in the said way. This result was part of von Neumann's attempts - now known to be flawed - to prove that \qm\ admits no hidden variables. }

Let $\CA$ and $\CB$ be \ca s, with (projective)\footnote{\label{tensorproducts}  The tensor product of two (or more) \ca s is not unique, and technically speaking we here need the so-called {\it projective} tensor product $\CA\hat{\ot}\CB$,  defined as the completion of the algebraic tensor product $\CA\ot\CB$  in the {\it maximal} $C^*$-cross-norm.
The choice of the projective tensor product guarantees that each state on
$\CA\ot\CB$ extends to a state on $\CA\hat{\ot}\CB$ by continuity; conversely,
since $\CA\ot\CB$ is dense in $\CA\hat{\ot}\CB$, each state on the latter is uniquely determined by its values on the former. See Takesaki (2003), Vol.\ {\sc i}, Ch.\ {\sc iv}.
In particular, product states $\rh\ot\sg$ and mixtures
$\om=\sum_i p_i \rh_i\ot\sg_i$ thereof as considered below are well defined on $\CA\hat{\ot}\CB$. If $\CA\subset \CB(\H_1)$ and $\CB\subset \CB(\H_2)$ are von Neumann algebras, as in the analysis of Raggio (1981, 1988), it is easier (and sufficient) to work with the {\it spatial} tensor product
$\CA\ovl{\ot}\CB$, defined as the double commutant (or weak completion)
of $\CA\ot\CB$ in $\CB(\H_1\ot \H_2)$. For any {\it normal} state on $\CA\ot\CB$ extends to a normal state on $\CA\ovl{\ot}\CB$ by continuity. Consequently,  Raggio's discussion is phrased in terms of {\it normal} states.
}  tensor product $\CA\hat{\ot}\CB$.
Less abstractly, just think of  $\CA=M_n(\C)$ and $\CB$ some (involutive) subalgebra of  $M_m(\C)$, such as the diagonal matrices $D_m(\C)$ mentioned above. The tensor product $\CA\hat{\ot}\CB$ is then the obvious subalgebra of $M_{nm}(\C)$, the algebra of all matrices on $\C^n\otimes C^m\cong \C^{nm}$.  In general, the interpretation of this setting is that $\CA$ and $\CB$ are the algebras of observables of two different physical systems, a priori quantum-mechanical in nature, but -  and this is the whole point - leaving open the possibility that one or both is {\it described} classically.  Indeed,
in our application to the Bohr--Einstein debate  $\CA$ is going to be the {\it noncommutative} algebra of observables of some quantum system, while $\CB$ will be  the {\it commutative} algebra of observables of the instrument.

Let us return to the case of general \ca s $\CA$ and $\CB$.
 A {\it product state} on $\CA\hat{\ot}\CB$ is a state of the form
$\om=\rh\ot\sg$, where the states $\rh$ on $\CA$ and $\sg$ on $\CB$ may be pure or mixed; the notation means that $\om(A\ot B)=\rh(A)\sg(B)$ for $A\in\CA$ and $B\in\CB$
(the value of $\om$ on more general elements of $\CA\hat{\ot}\CB$ then follows by linearity and - if necessary - continuity). 
We say that a state  $\om$ on $\CA\hat{\ot}\CB$
is {\it decomposable} or {\it classically correlated} when it is a mixture of product states, i.e.\ when
 $\om=\sum_i p_i \rh_i\ot\sg_i$, where the coefficients $p_i>0$ satisfy $\sum_i p_i=1$.\footnote{See Werner (1989). More precisely, $\om$ is decomposable if it is in the $w^*$-closure of the convex hull of the product states  on $\CA\hat{\otimes}\CB$.} A decomposable state $\om$ is pure precisely when it is a product of pure states,
 which is the case if $\om=\rh\ot\sg$ as above, but now  with both $\rh$ and $\sg$ pure. In this case
 - unlike for general pure states - the state $\om$  of the joint system is completely determined by  its restrictions $\rh=\om_{|\CA}$ and $\sg=\om_{|\CB}$ to $\CA$ and $\CB$, respectively.\footnote{This presupposes that $\CA$ and $\CB$ have units. The restriction $\om_{|\CA}$ of a state $\om$ on $\CA\hat{\otimes}\CB$ to $\CA$ is given by $\om_{|\CA}(A)=\om(A\ot 1)$, where $1$ is the unit element of $\CB$. Similarly, $\om_{|\CB}(B)=\om(1\ot B)$. } On the other hand, a state on $\CA\hat{\otimes}\CB$ may be said to be 
 {\it entangled} or  {\it \epr-correlated} (Primas, 1983)  when it is {\it not} decomposable.  An entangled {\it  pure} state has the property that its  restriction  to $\CA$ or $\CB$ is {\it mixed}.

Raggio's Theorem,\footnote{As adapted to \ca s (instead of von Neumann algebras) by Bacciagaluppi (1993).}  then, states that the following two conditions are equivalent:
\begin{itemize}{\it
\item Each state on $\CA\hat{\otimes}\CB$  is decomposable;
\item  $\CA$ or $\CB$ is commutative.}
\end{itemize}
 In other words, {\it \epr-correlated states exist  if and only if  $\CA$ and $\CB$ are both noncommutative.}

As one might expect, this result is closely related to the  Bell inequalities.\footnote{See the references
in footnote \ref{bellit} for references, which, however, do not contain Raggio's Theorem.}  
Consider  the CHSH-inequality (Clauser et al., 1969)
\beq \sup\{|\om(A_1(B_1+B_2)+A_2(B_1-B_2))|\} \leq 2, \label{bell}
\eeq
 where {\it for a fixed state $\om$} on  $\CA\hat{\otimes}\CB$ the supremum is taken over all self-adjoint operators
  $A_1,A_2\in \CA$, $B_1,B_2\in\CB$, each of norm $\leq 1$. It may then be shown that the two equivalent conditions just stated are in turn equivalent to a third one    (Baez, 1987; Raggio, 1988; Bacciagaluppi, 1993):\footnote{When $\CA$ and $\CB$ are both noncommutative, there surprisingly exist entangled mixed states that satisfy \er{bell};  the claim  that a state $\om$ satisfies \er{bell} whenever it is decomposable is valid only when $\om$ is pure (Werner, 1989; Werner \&\ Wolf, 2001; Seevinck, 2002). \L{wer}}
  \begin{itemize}{
\item   Each state $\om$ on $\CA\hat{\otimes}\CB$  satisfies \er{bell}.} 
\end{itemize}
Consequently, the inequality \er{bell} can only be violated in some (pure) state $\om$ when the algebras $\CA$ and $\CB$ are both noncommutative. If, on the other hand, \er{bell} is satisfied, then one knows that there exists a classical probability space and probability measure
(and hence a ``hidden variables" theory) reproducing the given correlations (Fine, 1982; Pitowsky, 1989). As stressed by  Bacciagaluppi (1993), such a description does {\it not} require the entire setting to be classical; as we have seen, only one of the algebras $\CA$ and $\CB$ has to be commutative for the  Bell inequalities to hold.

We are now in a position to understand the claim of our conciliatory
 ``Bohr". Suppose, as already indicated, that $\CA$ is the algebra of observables of some quantum system,
 and that $\CB$ is the algebra of observables of the instrument.
By definition of the word ``quantum", we suppose $\CA$ is noncommutative, as in the case $\CA=M_n(\C)$, whereas $\CB$ is commutative on  Bohr's doctrine of classical concepts.\footnote{If we regard $\CB$ as a commutative subalgebra of, say, the algebra of all observables in the Universe (or at least in some  laboratory), then different experimental contexts in the sense of Bohr are chosen by picking different such subalgebras. See Landsman (2005), \S 3.3 for a discussion of complementarity along these lines, and De Muynck (2004) for a competing recent account.}
We have now reached the fundamental point.
 By  Raggio's Theorem, {\it given the assumed noncommutativity of $\CA$}, the commutativity of $\CB$  is equivalent to the decomposability of all states of the joint system, which in turn is equivalent to the fact that
the restriction of each pure state of the joint system to either the observed system or the instrument is again pure. In other words, in Einstein's terminology each subsystem has its own ``real state", and this is precisely his {\it Trennungsprinzip}. 
The idea expressed here that Einstein requires states to be decomposable is reinforced if one accepts the usual arguments that Einstein's requirements lead to the Bell inequalities, since for pure states $\om$ the satisfaction of \er{bell} is equivalent to decomposability.\footnote{See footnote \ref{wer}.} 

Since all implications hold in both directions, we conclude that in physical theories whose observables are described by operators on a \Hs\ -  a class incorporating \qm\ as well as classical mechanics - {\it Bohr's doctrine of classical concepts} (construed as the need to
describe a given measuring device  by a commutative operator algebra) {\it
is mathematically equivalent to Einstein's separability principle, provided it is applied to the same measuring device in combination with the measured quantum system}.\footnote{This is not to say that
their - now joint -  doctrine is necessarily consistent. Indeed, most experts on the foundations of quantum theory would nowadays agree that the {\it classical} world is an {\it appearance} relative to the perspective of a certain class of observers, whereas the {\it quantum} world is {\it real} (though its peculiar  reality is  ``veiled"). See Landsman (2005) for a recent overview of this issue.
Provided that Nature is quantum-mechanical, classical physics is therefore deprived of its objective status altogether, undermining at least Bohr's reasoning. Einstein, on the other hand, could escape from this impasse by denying the premise.  Bohr's doctrine of classical conecpts has been largely endorsed  - only the emphasis on experiments being omitted - by the method of consistent histories (Omn\`{e}s, 1992; Gell-Mann \&\ Hartle, 1993; Griffiths, 2002). At the same time,  this approach  (together  with the Many Worlds Interpretation) provides the strongest indications that the
classical world is an appearance! It seems to follow that whereas Bohr's {\it doctrine} stands, its original {\it motivation} as being a requirement for objective science is questionable (Landsman, 2005).
De Muynck (2004) has drawn attention to a different reason why classical physics cannot be objective in the sense envisaged by Bohr and Heisenberg. He gives the example of a billiard ball, which in classical mechanics is seen as a rigid body.
The property of rigidity, however, is not objectively possessed but contextual, depending on the fact that under everyday conditions the  vibrations of the constituent molecules are negligible.
} This shows that Bohr's mechanism to gain objectification in \qm\ is mathematically equivalent to Einstein's.
\section{Who won?}\L{whowon}
To summarize our conclusions so far in a non-technical way, we can say that both Einstein and Bohr were realists of a subtle sort,  as follows:
\begin{itemize}{\it
\item Einstein's realism is the objectivity of spatially separated systems;
\item Bohr's realism is the objectivity of classical physics.}
\end{itemize}
In the context of \qm, then, these two special brands of realism actually {\it coincide}.

This analysis of objectification in quantum theory is seemingly unrelated to the
main theme of the Bohr--Einstein debate, viz.\ the (in)completeness of this theory. However, there is a direct connection between the two themes, as our analysis
enables us to give a version of Einstein's argument that \qm\ is incomplete as soon as it accommodates his separability principle, which - rationally speaking - Bohr would have had no choice but accepting. Namely,  if \qm\ is separable, then by Raggio's Theorem at least one of the two subsystems  which Einstein - on the basis of his criterion of objectivity - wishes  to separate from each other (by assigning each of them its own pure state), must be described classically. On the other hand, since Bohr would have been the first to agree that nothing physical could be said about the bare, uninterpreted mathematical formalism of quantum theory, his claim of completeness can only have related to the theory as interpreted through his doctrine of classical concepts.
Hence instead of directing his arrows at Heisenberg's uncertainty relations in his early attempts to prove the incompleteness of \qm, or his muddled later arguments based on multiple wave functions, Einstein's best bet would have been to simply tell  Bohr that he (Einstein) regarded a theory that necessarily describes part of the world classically {\it although the world as a whole is quantum-mechanical}, as incomplete. Indeed,
as we have seen in Section \ref{bohr}, it is precisely  this classical description that turns the bare theory - which is deterministic as it stands and could in principle have been endorsed by Einstein on this ground -  into the probabilistic one to which Einstein so famously objected that God would not have it. 

To state this argument in different words, let us  reconsider Einstein's remark (see Section \ref{einstein}) that `the [\epr] paradox forces us to relinquish one of the following two assertions:
\begin{trivlist}
\item [(1)] the description by means of the $\ps$-function is {\it complete}
\item[(2)] the real states of spatially separated objects are independent of each other.'
\end{trivlist}
Most physicists seem to agree that (1) - apparently Bohr's position - is right and (2) - Einstein's position - is wrong, leading to the conclusion that Bohr won the debate. However, this conclusion is superficial and preposterous, for Einstein's remark contains ``an essential ambiguity":  it is left unspecified whether it is meant to apply to either
\begin{trivlist}
\item[(i)]
the {\it bare mathematical theory} or 
\item[(ii)] 
 the {\it interpreted physical theory}. 
 \end{trivlist}
In the context of  the Bohr--Einstein debate, the only relevant interpretation of \qm\ is Bohr's, especially now that we have seen that his peculiar realism coincides with Einstein's. 
 \begin{trivlist}
\item[Ad (i).] In the first case, all current knowledge indicates that, indeed,  
among the two alternatives Einstein offers, (1) is right and (2) is wrong. 
Unfortunately for Bohr, this  case was of little interest to him (as all Bohr's  writings, particularly including his non-reply to the formal part of the \epr-argument, amply demonstrate). 
\item[Ad (ii).]  In the second case, one has precisely the opposite situation: 
a theory in which one has to restrict one's attention to a classical (i.e.\ commutative) subalgebra of the algebra of all (potential) ``observables" is manifestly incomplete,
whereas on the analysis in the preceding section this restricted theory is actually separable in the sense of Einstein. 
\end{trivlist}
 Our conclusion is that {\it on  the terms of the Bohr--Einstein debate}, it was 
{\it Einstein} who won. 

But how good is his victory from a broader perspective?
 One could certainly maintain that the  restriction to some classical subalgebra 
 renders a theory ``incomplete", but then:
\begin{quote}
 `who, learning that a theory is incomplete, could resist the idea that one ought to try to complete it?' (Fine, 1986, p.\ 88.)
 \end{quote}
Unfortunately for Einstein, the ``completion" of \qm\ adorned with the doctrine of classical concepts would simply be \qm\ itself: bare, uninterpreted, and \ldots nonseparable!
\section{The Talmudic philosopher}\label{maimonides}
\begin{quote}
`Yet, a certain difference in attitude and outlook remained (\ldots)'
(Bohr, 1949, p.\ 206.)
\end{quote}
Although the conditions {\it for} the acquisition of physical knowledge proposed by Einstein and Bohr turn out to be mathematically equivalent (in a world where observables are operators on a \Hs), they certainly disagreed about the status {\it of} this knowledge.
For while   Bohr insisted that the formalism of quantum  theory in principle provided a 
complete description of physics, he seems to have rejoiced in the {\it in}completeness
 of the {\it knowledge} this theory provides (i.e.\ upon application of his doctrine of classical concepts and its consequent probabilistic account of physics). 
 
  Spinoza - referring to the scholastic stance on the unknowability of God -  called this the `complacency of ignorance' (Donagan, 1996, p.\ 347); both the ignorance and the  complacency must have been unbearable to Einstein.
Thus  we have arrived at the true and insurmountable disagreement between Einstein and Bohr,  well captured by the latter's sneer to the effect that Einstein - famously claiming that God does not  play dice - should stop telling God what to do.\footnote{The Einstein claim is from his letter to Born of 4 December 1926, see Einstein \&\ Born (1969), pp.\ 129--130:
`Die Quantenmechanik ist sehr achtung-gebietend. Aber eine innere Stimme sagt mir, da\ss\ das noch nicht der wahre Jakob ist. Die Theorie liefert viel, aber dem Geheimnis des Alten bringt sie uns kaum n\"{a}her. Jedenfalls bin ich \"{u}berzeugt, da\ss\ der nicht w\"{u}rfelt.' The  Bohr quote  can be found on the web with high multiplicity but invariably without a reference.  A reliable source is Kroehling (1991), in which Pais says about Einstein that
`he had a certain type of arrogance. He had a certain belief that - not that he said it in those words but that is the way I read him personally - that he had a sort of special pipeline to God, you know. He would always say that God doesn't play dice to which Niels Bohr would reply ``but how do you know what God's doing?" He had these images of\ldots  that his notion of simplicity that that was the one that was going to prevail.' Here Bohr comes out more mildly than in the usual quotation as given in the main text. We are indebted to Michel Janssen for this source.}

Einstein's rejoinder is marvellous: in a letter to Schr\"{o}dinger from 19 June 1935 (discussed by  Howard (1985), Fine (1986), and Held (1998)), he portrays Bohr as follows:
\begin{quote}
`The Talmudic philosopher doesn't give a hoot for ``reality", which he regards as a hobgoblin of the naive (\ldots)' (Howard, 1985, p.\ 178; Howard's translation.)\footnote{ Here the original German is so delightful that we cannot resist quoting it: `Der taldmudistische Philosoph aber pfeift auf die ``Wirklichkeit" als auf einen Popanz der Naivit\"{a}t (\ldots)' Unfortunately,  Einstein's case rests on an equivocation (Held, 1998, \S 25).}
\end{quote}

Curiously, the few commentators on the Bohr--Einstein debate that perceive religious undertones in it (discussed by Jammer (1999) on pp.\
230--240) tend to put {\it Einstein} in the Talmudic tradition,\footnote{For example, commenting on Einstein (1936), Fokker (1955) wrote:  `His opinion culminates in a paradox: {\it Das ewig unbegreifliche an der Welt ist ihre Begreiflichkeit}. Our [mental] concepts are neither derived from experience, nor extracted from them, no, one does not have a relationship [between mental concepts and experience] as of {\it Suppe zum Rindfleisch, sondern eher wie die der Garderobenummer zum Mantel}. The concepts  we form are free creations of the mind. Einstein denies they can be derived from experience. I have always asked myself whether  Einstein's Jewish descent has played some role in this opinion of his. According to the  Old Testament God is the absolute other, with whom nothing can be compared, and of whom one accordingly is forbidden to form a picture. Within the tabernacle, the holiest of the holy, there is \ldots nothing. There is no thread that leads from the here, now, and us, to Him. The great wonder is that the absolute negation of us and the world, nonetheless interferes with the world and rules it. Would it be possible that this Old Testament notion of the great wonder has partly shaped Einstein's mind?'
(translated from the Dutch original by the present author).} leaving Bohr at the side of Eastern mysticism (a case supported by Bohr's choice of the yin-yang symbol as the emblem of his coat of arms following his Knighthood in 1947).

Far from analyzing such undertones here, we propose that the key difference between Bohr and Einstein could perhaps  be captured by a theological {\it analogy}.
This is undoubtedly all that Einstein himself had in mind in the above passage, and we merely wish to argue that {\it metaphorically} he was quite right in portraying Bohr in this way.
The analogy in question is between the knowability of Nature in physics, as limited by Bohr's doctrine of classical concepts, and the knowability of God in theology, highly restricted as  the Old Testament claims it to be. Indeed,  Bohr's idea that the quantum world can be studied exclusively through its influence on the ambient classical world has a striking parallel in the ``Talmudic" notion that God can only be known through his actions.    To illustrate this analogy, we  quote at some length from Maimonides's   famous {\it Guide of the Perplexed} from 1190:
\begin{quote}
`{\sc That} first and greatest of all thinkers, our teacher Moses, of blessed memory, made two requests and both his requests were granted. His first request was when he asked God to let him know His essence and nature; the second, which was the first in point of time, was when he asked Him to let him know His attributes. God's reply to the two requests was to promise that he would let him know all His attributes, telling him at the same time that they were His actions. Thereby He told him that His essence could not be apprehended in itself, but also pointed out to him a starting point from which he could set out to apprehend as much of Him as man can apprehend. And indeed, Moses apprehended more than anyone ever did before him or after him.

His request to know the attributes of the Lord is contained in the passage: {\it Shew me now thy ways and I shall know thee, to the end that I may find grace in thy sight} (Exodus 33, 13).  Consider carefully the wonderful expressions contained in this passage. The phrasing `Shew me now thy ways and I shall know thee' indicates that God is known by His attributes; if one knows the {\sc ways} one knows Him. (\ldots)

 After having requested the attributes of God, he asked for forgiveness for the people, and was granted forgiveness for them. Then he requested to apprehend God's essence, in the words {\it shew me now thy glory} (ibid.\ 18). Then only he was granted his first request, namely `shew me now thy ways', it being said to him: {\it I will make all my goodness pass before thee} (ibid.\ 19). The answer to the second request, however, was: {\it Thou canst not see my face: for there shall no man see me and live}   (ibid.\ 20).
 (\ldots)

 The outcome of our discussion is thus that the attributes which are applied to Him in Scripture are attributes of His acts, but He himself has no attributes.'
\\ (Maimonides, 1995, Book {\sc i}, Chapter {\sc liv}.)
\end{quote}

No direct influence of Maimonides on Bohr has been reported so far - unlike Einstein, Bohr was not well read in philosophy and theology - but it might be time to start looking for it;
Bohr was half Jewish. Einstein's intellectual inheritance from and admiration of Spinoza, on the other hand, is well documented; see Jammer (1999) for introductory remarks and Paty (1986) for a detailed account. Indeed, if Einstein had a hero at all, it may well  have been Spinoza.\footnote{Cf.\ the charming poem  `Zu Spinozas Ethik'  Einstein wrote in 1920:  `Wie lieb ich diesen edlen Mann
$\backslash\backslash$ Mehr als ich mit Worten sagen kann. $\backslash\backslash$ Doch f\"{u}rcht' ich, da\ss\ er bleibt allein $\backslash\backslash$ Mit seinem strahlenden Heiligenschein. (etc.)' See Jammer (1999), p. 267 for the complete text. Lorentz would be another candidate as Einstein's hero.}
Spinoza's opposition to Maimonides is abundantly clear from the following remarks:
\begin{quote}
 `The mind's highest good is the knowledge of God, and the mind's highest virtue is to know God.' (Spinoza, 1677, Part IV, Prop.\ 28.)\footnote{Translation by M.D. Wilson
 (Garrett, 1996, p.\ 90). This is not to say that Spinoza believed we can actually achieve complete knowledge of God.
Of God's infinitely many attributes we have access to only two, viz.\ extension and thought, all the other being hidden from us in principle. Furthermore, even God's two knowable attributes exist in infinitely many modi, whereas all human knowledge is neccesarily finite according to Spinoza.
}
 \end{quote}
\begin{quote}
`Since nothing can be conceived without God, it is certain that all those things which are in nature involve and express the concept of God, in proportion to their essence and perfection. Hence the more we cognize natural things, the greater and more perfect is the cognition of God we acquire, or, (since cognition of an effect through its cause is nothing but cognizing some property of that cause) the more we cognize natural things, the more perfectly do we cognize the essence of God, which is the cause of all things. so all our cognition, that is our greatest good, not only depends on the cognition of God but consists entirely in it.' (Spinoza, 1670, Ch.\ IV, \S 4.)\footnote{Translation by A. Donagan   (Garrett, 1996, p.\ 354). }
\end{quote}
According to  Donagan (1996, p.\ 347),  Spinoza `derided the medieval consensus [on having a very slight and inconsiderable knowledge of God] at a very early stage in his thinking', attributing their `complacency in ignorance'
to a fundamental philosophical mistake going back to Aristotle.\footnote{Namely the idea that legitimate definitions of substances must be by genus and difference, a mistake Spinoza thought Descartes had corrected by claiming that the  definition of a substance
 ought to be given by stating its attributes. Even within the Catholic tradition there had been disagreements on the knowability of God, in which e.g.\ Albertus Magnus (1200--1280) and Meister Eckhart (1260--1328) held positions comparable to Spinoza's - with the small but crucial difference that Spinoza (1632--1677) famously identified God with Nature ({\it Deus sive Natura}), whereas his predecessors talked about the Christian God. They were overruled, however, by St Thomas Aquinas (1224--1274), who held Maimonides's point of view.
See Kretzmann et al.\ (1988). } Similarly, it seems that Einstein
believed Bohr made a fundamental philosophical mistake somewhere,
although he  could not put his finger on the problem. 
\bigskip

\noindent {\bf Acknowledgement}
The author is indebted to
Michel Janssen for the invitation to write this paper, and to
Guido Bacciagaluppi, Jeremy Butterfield,
 Paul Juffermans, Willem de Muynck, and two anonymous referees for enlightening comments.
\section{References}
\begin{trivlist}
\item  Aspect, A., Grangier, P., \&\  Roger, G. (1981).
Experimental tests of realistic local theories via Bell's Theorem.
{\it  Physical Review Letters} 47, 460--467.
\item  Aspect, A., Grangier, P., \&\  Roger, G. (1982a). Experimental realization of Einstein-Podolsky-Rosen-Bohm Gedankenexperiment: A new violation of Bell's inequalities.
{\it Physical Review Letters} 49, 91--94.
\item Aspect, A.,  Dalibard, J., \&\  Roger, G. (1982b).
Experimental test of Bell's inequalities using time-varying analyzers.
{\it Physical Review Letters} 49, 1804--1807.
\item Aspect, A. (1992). Bell's theorem: the naive view of an experimentalist. In R.A. Bertlmann,  \&\ A. Zeilinger (Eds.), {\it Quantum [un]speakables} (pp.\ 119-153). Berlin:
Springer-Verlag.
\item Auletta, G. (2001). {\it Foundations and interpretation of quantum mechanics}. Singapore: World Scientific.
\item Bacciagaluppi, G. (1993). Separation theorems and Bell inequalities in algebraic quantum mechanics. In P. Busch, P.J. Lahti, \&\ P. Mittelstaedt (Eds.){\it Proceedings of the symposium on the foundations of modern physics
(Cologne, 1993)} (pp.\ 29--37). Singapore: World Scientific.
\item Baez, J. (1987).  Bell's inequality for $C\sp *$-algebras.
  {\it Letters in Mathematical Physics} 13,  135--136.
\item
Baggott, J.  (2004). {\it Beyond measure: Modern physics,
philosophy and the meaning of quantum theory}. Oxford: Oxford
University Press.
\item  Bell, J.S. (1987). {\it Speakable and unspeakable in quantum mechanics}. Second Edition.
Cambridge: Cambridge University Press.
\item  Bell, J.S. (2001).  {\it John S. Bell on the foundations of quantum mechanics}.
Singapore: World Scientific.
\item  Beller, M. (1999). {\it Quantum dialogue}.  Chicago: University of Chicago Press.
\item  Beller, M. \&\ Fine, A. (1994). Bohr's response to \epr.
In J. Faye, \&\ H. Folse (Eds.) {\it Niels Bohr and contemporary
philosophy} (pp.\ 1--31).
  Dordrecht: Kluwer Academic
Publishers.
\item Bennett, C.H.,  Brassard, G.,  Crepeau, C., Jozsa, R.,  Peres, A., \&\ Wootters, W. (1993). Teleporting an unknown quantum state via dual classical and
 \epr\ channels. {\it Physical Review Letters} 70, 1895--1899.
\item Bohr, N. (1927) The quantum postulate and the recent development of atomic theory. {\it Atti del Congresso Internazionale dei Fisici (Como, 1927)}.
Reprinted in Bohr (1934) and in Bohr (1985).
\item Bohr, N. (1934). {\it Atomic theory and the description of nature}.
Cambridge: Cambridge University Press.
\item Bohr, N. (1935). Can quantum-mechanical description of physical reality be considered complete? {\it Physical Review} 48, 696--702.
\item Bohr, N. (1937). Causality and complementarity. {\it Philosophy of Science} 4, 289--298.
\item Bohr, N. (1949). Discussion with Einstein on epistemological problems in atomic physics. In P.A. Schilpp (Ed.) {\it Albert Einstein: Philosopher-scientist} (pp.\ 201--241). La Salle: Open Court.
\item Bohr, N. (1956). Mathematics and natural philosophy. {\it Scientific Monthly} 82, 85--88.
\item Bohr, N. (1958). {\it Atomic physics and human knowledge}. New York: Wiley.
\item Bohr, N. (1985). {\it Collected works. Vol.\ 6: Foundations of quantum physics {\sc i} (1926--1932)}. Kalckar, J. (Ed.). Amsterdam: North-Holland.
\item Bohr, N. (1987). {\it The philosophical writings of Niels Bohr}. 3 Vols.
Woodbrigde: Ox Bow Press.
\item Bohr, N. (1996). {\it Collected works. Vol.\ 7: Foundations of quantum physics {\sc ii} (1933--1958)}. Kalckar, J. (Ed.). Amsterdam: North-Holland.
\item Bub, J. (1974). {\it The interpretation of quantum mechanics}. Dordrecht: D. Reidel.
\item Bub, J. (1997). {\it Interpreting the quantum world}.  Cambridge: Cambridge University Press.
\item Bub, J. (2004). Why the quantum? {\it Studies in History and Philosophy of Modern Physics } 35, 241--266.
\item Camilleri, K. (2005). {\it Heisenberg and quantum mechanics: The evolution of a philosophy of nature}. Ph.D.\  Thesis, University of Melbourne.
\item Clauser, J.F., Horne, M.A., Shimony, A., \&\ Holt, R.A. (1969).  Proposed experiment to test local hidden-variable theories. {\it  Physical Review Letters} 23, 880--884
\item  Clifton, R.,  Bub, J., \&\ Halvorson, H. (2003).
Characterizing quantum theory in terms of information-theoretic constraints.
 {\it Foundations of Physics} 33, 1561--1592.
\item Cushing, J.T. (1994). {\it Quantum mechanics: Historical contingency and the Copenhagen hegemony}.   Chicago: University of Chicago Press.
\item Daniel, W. (1989).
Bohr, Einstein and realism. {\it Dialectica}  43, 249--261.
\item De Muynck, W.M. (2004). Towards a Neo-Copenhagen interpretation of quantum mechanics.
{\it Foundations of Physics}  34, 717--770.
\item  Deltete, R. \&\ Guy, R. (1991). Einstein and \epr.
{\it  Philosophy of Science} 58, 377--397.
\item Dickson, M. (2002). Bohr on Bell: a proposed reading of Bohr and its implications for Bell's theorem. In T. Placek, \&\ J. Butterfield (Eds.).  {\it Non-locality and modality}   (pp.\ 19--36).  Dordrecht: Kluwer Academic Publishers.
\item Donagan, A. (1996). Spinoza's theology. In Garrett (1996), pp.\ 343--382.
\item  Einstein, A. (1930). Mein Glaubensbekenntnis. In K. Sander (Ed.) {\it Albert Einstein: Verehrte An- und Abwesende!
Originaltonaufnahmen 1921--1951}.  2-CD-Set plus Booklet (ISBN
3-932513-44-4). K\"{o}ln: suppos\'{e}.\footnote{For those who do
not possess these CD's, the quotation in footnote \ref{live} can
be heard in Einstein's own voice  from
\texttt{http://www.suppose.de/texte/einstein.html} by clicking on
\texttt{H\"{o}rprobe}.}
\item Einstein, A. (1936). Physik und Realit\"{a}t. {\it Journal  of the Franklin Institute} 221, 313--347. English translation in Einstein (1956), pp.\ 59--97.
\item Einstein, A. (1948). Quanten-Mechanik und Wirklichkeit. {\it Dialectica} 2, 320--324.
\item Einstein, A. (1949a). Autobiographical notes. In P.A. Schilpp (Ed.){\it Albert Einstein: Philosopher-Scientist} (pp.\ 1--94). La Salle: Open Court.
\item Einstein, A. (1949b). Remarks to the essays appearing in this collective volume.
(Reply to criticisms). In P.A. Schilpp (Ed.)
 {\it Albert Einstein: Philosopher-scientist} (pp.\ 663--688). La Salle: Open Court.
 \item Einstein, A. (1954). {\it Ideas and opinions}. New York: Bonanza Books.
  \item Einstein, A. (1956). {\it Out of my later years}. Secaucus (N.J.): Citadel Press.
  \item Einstein, A. (1982). {\it Mein Weltbild}. Frankfurt: Ullstein.
\item Einstein, A. \&\ Born, M. (1969). {\it Briefwechsel 1916--1955}. M\"{u}nchen: Nymphenburger Verlagshandlung.
\item Einstein, A., Podolsky, B., \&\ Rosen, N. (1935). Can quantum-mechanical description of physical reality be considered complete? {\it Physical Review} 47, 777--780.
\item ÊEmch, G.G. (1984) {\it Mathematical and conceptual foundations of 20th-Century Physics}. Amsterdam: North-Holland.
\item   Faye, J. (1991). {\it Niels Bohr: His heritage and legacy. An anti-realist view of quantum mechanics}. Dordrecht: Kluwer Academic Publishers.
\item Fine, A. (1982). Joint distributions, quantum correlations, and commuting observables.
{\it Journal of Mathematical Physics} 23, 1306--1310.
\item Fine, A. (1986). {\it The shaky game}. Chicago: University of Chicago Press.
\item Fine, A. (2004). The Einstein-Podolsky-Rosen argument in quantum
theory. In E.N. Zalta, (Ed.) {\it The Stanford Encyclopedia of
Philosophy (Spring 2004 Edition)}.
\\\texttt{http://plato.stanford.edu/archives/sum2004/entries/qt-epr/}.
\item Fokker, A.D. (1955). Albert Einstein. 14 maart 1878 (sic) - 18 april 1955.
{\it Nederlands Tijdschrift voor Natuurkunde} mei, 125--129. In Dutch. Reprinted in
{\it ibid} april 2005, 104--106.
\item
Folse, H.J. (1985). {\it  The philosophy of Niels Bohr}.
Amsterdam: North-Holland.
\item  French, A.P., \&\   Kennedy, P.J. (Eds.). (1985). {\it Niels Bohr: A centenary volume}. Cambridge (MA): Harvard University Press.
 \item Garrett, D. (Ed.). (1996). {\it The Cambridge companion to Spinoza}.
Cambridge: Cambridge University Press.
\item Gell-Mann, M. (1979). Is nature simple? In D. Huff, \&\ O. Prewett, (Eds.){\it The nature of the physical universe, 1976 Nobel Conference},
(p.\ 29). New York: Wiley Interscience.
\item  Gell-Mann, M. \&\ Hartle, J.B. (1993). Classical equations for quantum systems.
 {\it Physical Review} D47, 3345--3382.
\item Gill, R.D. \&\  Larsson, J.-A. (2004). Bell's inequality and the coincidence time loophole. {\it Europhysics Letters} 67, 707--713.
\item  Gill, R.D.,  Weihs, G.,  Zeilinger, A., \&\  Zukowski, M.  (2002).
 No time loophole in Bell's theorem: The Hess--Philipp model is nonlocal.
{\it Proceedings of the National Academy of Science} 99, 14632--14635.
\item  Gill, R.D.,  Weihs, G.,  Zeilinger, A., \&\  Zukowski, M.  (2003).
 Comment on `Exclusion of time in the theorem of Bell' by K. Hess and W. Philipp. {\it
 Europhysics  Letters} 61, 282--283.
\item Grangier, P. (2001). Count them all. {\it Nature} 409, 774--775.
\item
Greenberger, D.M., Horne, M.A., Shimony, A., \&\ Zeilinger A. (1990).
Bell's theorem without inequalities. {\it  American Journal of Physics}  58,  1131--1143.
\item  Griffiths, R.B. (2002).  {\it   Consistent quantum theory}.
Cambridge: Cambridge University Press.
\item   Haag, R. (1992).   {\it Local quantum physics: Fields, particles, algebras}.
Heidelberg: Springer-Verlag.
\item Halvorson, H. \&\ Clifton, R.  (2002). Reconsidering Bohr's reply to \epr. In T. Placek, \&\ J. Butterfield (Eds) . {\it Non-locality and modality} (pp.\ 3--18).
Dordrecht: Kluwer Academic Publishers.
\item Hardy, L. (1993). Nonlocality for two particles without inequalities for almost all entangled states. {\it  Physical Review Letters} 71, 1665--1668.
\item Heisenberg, W. (1958). {\it Physics and philosophy: The revolution in modern science}. London: Allen \&\ Unwin.
\item Held, C. (1994). The meaning of complementarity. {\it Studies in History and Philosophy of Science} 25, 871--893.
\item Held, C. (1998). {\it Die Bohr-Einstein-Debatte: Quantenmechanik und Physikalische Wirklichkeit}.  Paderborn: Sch\"{o}ningh.
\item Hendry, J. (1984). {\it The creation of quantum mechanics and the Bohr-Pauli Dialogue}.  Dordrecht: D. Reidel.
\item Hess, K.  \&\ Phillip, W. (2001a). A possible loophole  in the theorem of Bell.
{\it Proceedings of the National Academy of Science}
98, 14244--14227.
\item Hess, K.  \&\ Phillip, W. (2001b).  Bell's theorem and the problem
 of decidability between the views of Einstein and Bohr.
{\it Proceedings of the National Academy of Science}
98,  14228--14233.
\item Hess, K.  \&\ Phillip, W. (2002).  Exclusion of time in the theorem of Bell.   {\it
 Europhysics  Letters} 57, 775--781.
\item Hess, K.  \&\ Phillip, W. (2004). Breakdown of Bell's theorem for certain objective local parameter spaces. {\it Proceedings of The National Academy of Science} 101, 1799--1805.
\item Hess, K.  \&\ Phillip, W. (2005). The Bell theorem as a special case of a theorem of Bass. Preprint.
\item Honner, J. (1987). {\it The description of nature: Niels Bohr and the philosophy of quantum physics}. Oxford: Oxford University Press.
\item Hooker, C.A. (1972). The nature of quantum mechanical reality: Einstein versus
Bohr. In J. Colodny, (Ed.) {\it Paradigms \&\ paradoxes: The
philosophical challenges of the quantum domain} (pp.\ 67--302).
Pittsburgh: University of Pittsburgh Press.
\item Hooker, C.A. (1991). Projection, physical intelligibility, objectivity and completeness: the divergent ideals of Bohr and Einstein. {\it The British Journal for the Philosophy of Science} 42, 491--511.
\item  Howard, D. (1985). Einstein on locality and separability. {\it Studies in History and Philosophy of Science} 16, 171--201.
\item Howard, D. (1990). `Nicht sein kann was nicht sein darf', or the prehistory of \epr, 1909-1935: Einstein's early worries about the \qm\ of composite systems. In A.I. Miller (Ed.){\it
 Sixty-two years of uncertainty} (pp.\ 61--11). New
 York: Plenum.
\item Howard, D. (1994). What makes a classical concept classical? Towards a reconstruction of Niels Bohr's philosophy of physics. In J. Faye, \&\ H. Folse (Eds.){\it Niels Bohr and contemporary philosophy}
(pp.\ 201--229).  Dordrecht: Kluwer Academic Publishers.
\item Howard, D. (2004a). Who invented the Copenhagen interpretation?
{\it Philosophy of Science} 71, 669-682.
\item  Howard, D. (2004b). Einstein's philosophy of science. In E.N. Zalta, (Ed.).{\it
The Stanford Encyclopedia of Philosophy (Spring 2004 Edition)}.
\\
\texttt{http://plato.stanford.edu/archives/spr2004/entries/einstein-philscience/}.
\item Howard, D. (2005). Einstein's philosophy of science. In M. Janssen, \&\ C. Lehner (Eds.){\it The Cambridge companion to Einstein}. In press. New York: Cambridge University Press.
\item Israel, J. (1995). {\it The Dutch republic: Its rise, greatness, and fall 1477-1806}.
Oxford: Oxford University Press.
\item Jammer, M. (1974). {\it The philosophy of quantum mechanics}.  New York: Wiley.
 \item Jammer, M. (1999). {\it Einstein and religion: Physics and theology}. Princeton:
Princeton University Press.
\item Jarrett, J.P. (1984).  On the physical significance of the locality conditions in the Bell arguments. {\it  No\^{u}s} 18, 569-589.
\item  Kretzmann, N., Kenny, A., Pinborg, J., \&\ Stump, E. (1988). {\it
The Cambridge history of later medieval philosophy}. Cambridge:
Cambridge University Press.
\item Krips, H.P. (1969). Two paradoxes in \qm. {\it Philosophy of Science} 36, 145--152.
\item Kroehling, R. (Director) (1991). {\it Albert Einstein: How I see the world}.  PBS Home Video.
\item Landsman, N.P. (1998). {\it Mathematical topics between classical and quantum mechanics}. New York: Springer-Verlag.
 \item Landsman, N.P. (2005). Between classical and quantum. In J. Earman, \&\ J.
 Butterfield (Eds.)
{\it  Handbook of the philosophy of science, Vol.\ 2:  Philosophy
of physics}, in press. Amsterdam: Elsevier.
\texttt{arXiv:quant-ph/0506082}.
\item Maimonides, M. (1995). {\it The Guide of the perplexed}. Guttmann, J. (Ed.).
Indianapolis: Hackett.
\item McGrath, J.H. (1978).  A formal statement on the Einstein--Podolsky--Rosen argument.
{\it   International Journal of Theoretical Physics } 17, 557--571.
\item  Mehra, J. \&\ and Rechenberg, H. (2001). {\it The historical development of quantum theory. Vol.\ 6: The completion of quantum mechanics 1926--1941.    Part 2: The conceptual completion of quantum mechanics.}  New York: Springer-Verlag.
\item Mermin, N.D. (1993). Hidden variables and the two theorems of John Bell.
{\it Reviews of Modern Physics} 65, 803--815.
\item Murdoch, D. (1987). {\it Niels Bohr's philosophy of physics}. Cambridge: Cambridge University Press.
\item Myrvold, W.C. (2003). A loophole in Bell's theorem? Parameter dependence in the Hess-Philipp model. {\it Philosophy of Science}  70, 1357--1367.
\itemÊOmn\`{e}s, R.  (1992). Consistent interpretations of quantum mechanics. {\it  Reviews of Modern Physics }  64, 339--382.
\item Pais, A. (1982). {\it Subtle is the lord: The science and life of Albert Einstein}. Oxford: Oxford University Press.
\item Pais, A. (1991). {\it
Niels Bohr's times: In physics, philosophy, and polity}.  Oxford:
Oxford University Press.
\item Pais, A. (2000). {\it The genius of science}.  Oxford: Oxford University Press.
\item Paty, M. (1986). Einstein and Spinoza. In M. Grene, \&\ D. Nails (Eds.) {\it Spinoza and the sciences} (pp.\ 267--302).
. Dordrecht: Kluwer.
\item Peres, A. (2002). Karl Popper and the Copenhagen interpretation.
 {\it Studies in History and Philosophy of Modern Physics } 33, 23--34.
\item Pitowsky, I. (1989). {\it Quantum probability - quantum logic}. Berlin: Springer.
\item Primas, H. (1983). {\it Chemistry, quantum mechanics and reductionism}. Second Edition. Berlin: Springer-Verlag.
\item Raggio, G.A. (1981). {\it States and composite systems in $W^*$-algebras quantum mechanics}. Ph.D Thesis, ETH Z\"{u}rich.
\item Raggio, G.A. (1988).
  A remark on Bell's inequality and decomposable normal states.
  {\it Letters in Mathematical Physics}  15, 27--29.
\item Redhead, M.L. (1987). {\it Incompleteness, nonlocality and realism: A prolegomenon to the philosophy of quantum mechanics}. Oxford: Clarendon Press.
\item Rosenfeld, L. (1967). Niels Bohr in the thirties. Consolidation and extension of the conception of complementarity. In S. Rozental (Ed.) {\it Niels Bohr: His life and work as seen by his friends and colleagues} (pp.\ 114--136). Amsterdam: North-Holland.
\item Santos, E. (2005). Bell's theorem and the experiments: Increasing empirical support for local realism?
 {\it Studies in History and Philosophy of Modern Physics} 36, 544--565.
\item Scheibe, E. (1973). {\it The logical analysis of quantum mechanics}. Oxford: Pergamon Press.
\item Scheibe, E. (2001).
{\it Between rationalism and empiricism: Selected papers in the
philosophy of physics}. New York: Springer-Verlag.
\item Seevinck, M. (2002). {\it
Entanglement, local hidden variables and Bell-inequalities: An
investigation in multi-partite quantum mechanics.} M.Sc.\ Thesis,
University of Utrecht.\\
\texttt{http://www.phys.uu.nl/~wwwgrnsl/seevinck/scriptie.pdf}.
\item Seevinck, M. \&\ Uffink, J. (2002). Sufficient conditions for three-particle entanglement and their tests in recent experiments. {\it Physical Review}  A65, 012107.
\item  Shimony, A. (2001). The logic of \epr. {\it Annales de la Fondation Louis de Broglie} 26 (no sp\'{e}cial 3/3), 399--410.
\item Shimony, A. (2005). Bell's Theorem. In E.N. Zalta, (Ed.)
 {\it
The Stanford Encyclopedia of Philosophy (Spring 2004 Edition)}.
\texttt{http://plato.stanford.edu/archives/sum2005/entries/bell-theorem/}.
\item Spinoza, B. de (1670). {\it Tractatus theologico-politicus}.
Hamburg (Amsterdam).
\item Spinoza, B. de (1677). {\it Ethica}. Amsterdam.
\item Szabo, L. \&\ Fine, A. (2002).  A local hidden variable theory for the GHZ experiment. {\it Physics Letters} A295, 229--240.
\item Takesaki, M. (2003). {\it Theory of operator algebras.}   Vols.\ I-III. New York: Springer-Verlag.
\item  Tittel, W.,  Brendel, J.,  Zbinden, H., \&\  Gisin, N. (1998).
Violation of Bell inequalities by photons more than 10 km apart.
{\it  Physical Review Letters}  81, 3563--3566.
\item Ursin, R.,  Jennewein, T.,  Aspelmeyer, M., Kaltenbaek, R., Lindenthal, M., Walther, P., \&\ Zeilinger, A. (2004).
Quantum teleportation across the Danube. {\it Nature} 430, 849.
\item Werner, R.F. (1989). Quantum states with Einstein-Podolsky-Rosen correlations admitting a hidden-variable model. {\it Physical Review} A40, 4277Ð4281.
\item Werner, R.F. \&\ Wolf, M.M. (2001). Bell inequalities and entanglement. \texttt{arXiv.org/quant-ph/0107093}.
\item Wheeler, J.A. (1985).  Physics in Copenhagen in 1934 and 1935.
In French \&\ Kennedy (1985),  pp.\ 221--226.
\item Whitaker, A. (1996). {\it Einstein, Bohr, and the quantum dilemma}. Cambridge: Cambridge University Press.
\item Whitaker, M.A.B. (2004). The \epr\ paper and Bohr's response: a re-assessment. {\it Foundations of Physics}
34, 1305--1340.
\item Winsberg, E., \&\ Fine, A. (2003). Quantum life: interaction, entanglement and separation. {\it Journal of Philosophy} 100, 80--97.
\item Wood, A. (1954). {\it Thomas Young: Natural philosopher 1773--1829}.
Cambridge: At the University Press.
\item  Zeilinger, A. (2000). Quantum teleportation. {\it Scientific American} April, 32--41.
\end{trivlist}
\end{document}